\documentstyle[aps,prd,epsf]{revtex}

\newcommand{\be}{\begin{equation}}
\newcommand{\ee}{\end{equation}}
\newcommand{\bea}{\begin{eqnarray}}
\newcommand{\eea}{\end{eqnarray}}
\newcommand{\bdm}{\begin{displaymath}}
\newcommand{\edm}{\end{displaymath}}

\newcommand{\diff}{\mbox{$\rm{d}$}}
\newcommand{\Diff}{\mbox{$\rm{D}$}}
\newcommand{\hatDiff}{\hat{\Diff}}

\newcommand{\fourA}{A^{(4)}}

\newcommand{\we}{\wedge}
\newcommand{\sprod}[2]{\langle #1\, , \,#2 \rangle}
\newcommand{\Trace}[1]{\mbox{Tr} \left\{ #1 \right\}}

\newcommand{\taur}{\tau_{r}}
\newcommand{\tautheta}{\tau_{\vartheta}}
\newcommand{\tauphi}{\tau_{\varphi}}
\newcommand{\hatast}{\hat{\ast}}
\newcommand{\hatlap}{\hat{\Delta}}
\newcommand{\Ybold}{\mbox{\boldmath $Y$}}
\newcommand{\YLJM}{{\Ybold}^{L}_{JM}}
\newcommand{\Yarg}[1]{{\Ybold}^{#1}_{JM}}
\newcommand{\CLJM}{C^{L}_{JM}}
\newcommand{\Carg}[1]{C^{#1}_{JM}}
\newcommand{\sigbold}{\mbox{\boldmath $\sigma$}}
\newcommand{\taubold}{\mbox{\boldmath $\tau$}}
\newcommand{\radbold}{\mbox{\boldmath $r$}}

\begin{document}

\title{On rotational excitations and axial deformations of 
       BPS monopoles and Julia-Zee dyons}

\author{M. Heusler, N. Straumann, and M. Volkov}

\address{Institute for Theoretical Physics \\
         The University of Zurich \\
         CH--8057 Zurich, Switzerland}

\date{\today}

\maketitle

\begin{abstract}

It is shown that Julia-Zee dyons do not admit slowly 
rotating excitations. This is achieved by investigating 
the complete set of stationary excitations which can 
give rise to non-vanishing angular momentum. The relevant
zero modes are parametrized in a gauge invariant way and 
analyzed by means of a harmonic decomposition. Since 
general arguments show that the solutions to the 
linearized Bogomol'nyi equations cannot contribute to 
the angular momentum, the relevant modes are governed by 
a set of electric and a set of non self-dual magnetic 
perturbation equations. The absence of axial dipole 
deformations is also established.

\end{abstract}


\section{Introduction}
\label{section-I}

The main question addressed in this paper is whether 
Julia-Zee dyons admit rotational excitations. The 
investigation of this problem was motivated by some 
surprising results which we recently obtained for a class of 
self-gravitating non-Abelian soliton and black hole 
configurations. In \cite{BHSV97} we showed that the 
Bartnik-McKinnon solutions \cite{Bartnik88} admit slowly 
rotating excitations. A two-parameter family of axisymmetric
excitations of the static black hole solutions to the 
Einstein-Yang-Mills system was established as well. In 
addition to the charged, rotating black holes found in 
\cite{Volkov97}, there also exists a branch of uncharged,
rotating black holes, as well as a branch of 
stationary -- but not static -- black holes with vanishing 
Komar angular momentum \cite{BHSV97}.

On the other hand, the situation was shown to be completely 
different in the presence of scalar fields \cite{BH97}. 
Slowly rotating generalizations of (self-gravitating)
{\it solitons\/} were {\it excluded\/} for a relatively 
large class of theories with non-Abelian gauge fields 
coupled to Higgs fields. In particular, the results obtained
in \cite{BH97} apply to the t'Hooft-Polyakov monopole and its
self-gravitating generalizations. For {\it black hole\/} 
solutions of gauge theories with Higgs fields the situation 
is again different: Rotating excitations of static black 
holes generically exist; they are, however, necessarily 
charged.

Since we are still lacking a deeper physical understanding 
of the facts mentioned above, we have been looking 
for other (not gravitating) examples which might help to 
find a clue. On the basis of our previous experience, we 
originally expected static {\it dyon\/} solutions to admit 
rotational excitations. The simplest examples are the 
Julia-Zee dyons, which are related to the 
Bogomol'nyi-Prasad-Sommerfield (BPS) monopole by a 
one-parameter family of hyperbolic rotations in internal 
space. 

The problem of small fluctuations around BPS monopoles has 
been examined some time ago by Mottola \cite{Mottola78}, 
Adler \cite{Adler79}, Weinberg \cite{Weinberg79}, and 
completed in a comprehensive analysis by 
Akhoury {\it et.al.\/} \cite{Akhoury80}. The main emphasis 
was placed on the study of normalizable zero modes in the 
{\it self-dual\/} sector, because these are relevant to the 
structure of multi-monopole solutions. The moduli space of 
SU(2) monopole solutions carrying $n$ units of magnetic 
charge was shown to be $4n$-dimensional \cite{Weinberg79}, 
\cite{Weinberg80}. (See also \cite{Manton82}, 
\cite{GibbonsManton95}, and \cite{Lee-etal96} for a 
generalization to arbitrary gauge groups and for further 
references.) As these studies are dealing with the self-dual
sector, an investigation of the remaining zero modes seems 
to be necessary. This is also motivated by the following 
observations, which are obtained from general considerations.
\begin{itemize}
\item
The solutions to the linearized Bogomol'nyi 
equations -- independently of whether or not they are 
physically acceptable -- cannot give rise to a non-vanishing 
angular momentum. This is true for both BPS monopoles and 
Julia-Zee dyons.
\item
The only excitations of BPS monopoles which can contribute 
to the angular momentum arise from perturbations of the time
component, $\delta \Phi \equiv \delta A_{t}$, of the gauge 
potential; these will be called {\it electric\/} modes.
\item
The only perturbations of Julia-Zee dyons which can 
contribute to the angular momentum are the electric ones 
and the non self-dual magnetic ones. (A mode will be
called {\it magnetic\/}, if $\delta \Phi$ vanishes, and 
{\it non self-dual\/}, if it is a solution of the linearized
field equations, but not of the linearized Bogomol'nyi 
equations.)
\end{itemize}

The full problem, including the non self-dual fluctuations, 
was studied by Baake \cite{Baake92} in connection with the 
stability analysis of the t'Hooft-Polyakov monopole. In his 
work, Baake mainly focused on the {\it negative} fluctuation
modes, the absence of which he was able to prove by applying
the Jacoby criterion. Since we are not aware of any other 
work devoted to non self-dual zero modes, we carry out a 
systematic, gauge invariant perturbation analysis in order 
to study the rotational excitations of BPS monopoles and 
Julia-Zee dyons. The emphasis in the present paper is mainly
placed on the methods. The main result is, 
unfortunately, negative: Neither BPS monopoles nor 
Julia-Zee dyons admit slowly rotating excitations. 

A further motivation for studying non self-dual rotational 
excitations is provided by a theorem due to Taubes 
\cite{Taubes82}, according to which not every finite energy 
solution to the field equations in the BPS limit has to 
satisfy the {\it first order\/} Bogomol'nyi equations. 
Hence, the existence of physically acceptable excitations 
orthogonal to the Bogomol'nyi sector is not a priori 
excluded. However, the results of the present work imply 
that all non self-dual axisymmetric finite energy solutions,
if they exist, are necessarily {\it disconnected\/} from the
Julia-Zee dyons. This is, in fact, a week version of the 
original conjecture \cite{Jaffe80} (the general form of
which was disproved in Taubes' work \cite{Taubes82}).
It is, however, likely that configurations with unit 
winding number and discrete angular momenta exist. This 
is, for instance, the case for boson stars \cite{BOSONSTAR}. 

This paper is organized as follows: 
In Sect. \ref{section-II} we briefly review the symmetry 
which connects the PBS monopole solution with the 
one-parameter family of Julia-Zee dyons. In 
Sect. \ref{section-III} we show how to use this symmetry
to reduce the perturbation analysis for Julia-Zee dyons to 
that for the PBS monopole. The main advantage of this consists 
in the fact that, after a hyperbolic rotation, the electric 
background field vanishes. This implies that -- in the
rotated system -- the electric perturbations, $\delta \Phi$,
do not couple to the magnetic ones. We then show that the 
non self-dual magnetic perturbations are governed by a 
system of first order equations for a one-form, $\delta B$.
The latter comprises the perturbations of the Higgs field, 
$\delta H$, and the perturbation of the three-dimensional 
gauge potential, $\delta A$, in a gauge invariant way. 

In Sect. \ref{section-IV} we present the decomposition of 
the gauge invariant perturbations $\delta \Phi$ and 
$\delta B$ in terms of isospin harmonics. We also show that 
the expression for the angular momentum can be integrated, 
implying that only the boundary values of the perturbation 
amplitudes are relevant. The complete set of perturbation 
equations is derived in Sect. \ref{section-V}. 
This consists of an even and an odd parity sector. 
Each sector comprises the 
electric equations for $\delta \Phi$, the magnetic equations
for $\delta B$ (governing the non self-dual modes), and the
inhomogeneous linearized Bogomol'nyi equations for 
$\delta H$ and $\delta A$ in terms of the source $\delta B$.

In Sect. \ref{section-VI} 
we discuss the odd parity perturbations and 
present the solutions of the complete set of equations in 
closed form. As the odd parity modes cannot contribute to 
the angular momentum, we conclude from the solutions that
there exist no physically acceptable axial dipole
{\it deformations\/} of Julia-Zee dyons. The more 
interesting {\it even\/} parity modes are discussed in 
Sect. \ref{section-VII}. 
We show how to use the explicitly known 
solutions to reduce the {\it magnetic\/} 
problem to a standard Schr\"odinger equation. 
We also prove that the {\it electric\/} 
perturbations are governed by exactly the same equation. 
Since the latter has a non-negative potential, we are able 
to present a rigorous discussion of all modes. It turns out 
that there exist solutions (both electric and magnetic)
which give rise to finite angular momentum. However, none
of these modes are regular.

\section{BPS monopoles and Julia-Zee dyons}
\label{section-II}

We consider stationary solutions to the SU(2) 
Yang-Mills-Higgs (YMH) equations with gauge potential 
$\fourA$ and Higgs triplet $H$ in the BPS limit (i.e., 
without Higgs self-interaction). The 
dimensionally reduced YMH action becomes 
\be
S = \frac{1}{2} \int \left\{ (F,F) + (\Diff H,\Diff H) -
(\Diff \Phi,\Diff \Phi)-[\Phi,H]^2 \right\} \diff ^3 x ,
\label{action-1}
\ee 
where $\Phi$ and $A$ parametrize the electric and the magnetic 
components of the gauge potential,
\be
\fourA = \Phi \, \diff t + A .
\label{pot-1}
\ee
The quantities $F$ and $\Diff$ denote the field strength 
two-form and the gauge covariant derivative with respect to 
the three-dimensional magnetic potential $A$:
\be
F= \diff A +  A \we A, \ \ \ 
\Diff \Phi = \diff \Phi + [A,\Phi], \ \ \ 
\Diff H    = \diff H + [A,H].
\label{pot-2}
\ee
(For arbitrary Lie algebra valued $p$-forms $\alpha$ the
inner product is defined by 
$(\alpha,\alpha) \diff ^{3} x = 
\Trace{\alpha \we \ast \alpha}$, where $\ast$ is the 
three-dimensional Hodge dual.)

The perturbation analysis for Julia-Zee (JZ) dyons will be
simplified considerably by the fact that the dimensionally 
reduced action (\ref{action-1}) is invariant under 
hyperbolic rotations in the $(H , \Phi)$-plane; that is, 
the transformation 
\be
\left(\begin{array}{c}
H \\ \Phi
\end{array}\right)
\rightarrow
\left(\begin{array}{cc}
\cosh(\gamma) & \sinh(\gamma) \\
\sinh(\gamma) & \cosh(\gamma)
\end{array}\right)
\left(\begin{array}{c}
H \\ \Phi 
\end{array} \right)
\label{boost}
\ee
is a symmetry of the action (\ref{action-1}).

In particular, a BPS monopole solution $H = H_{\rm mon}$, 
$\Phi = 0$ with magnetic charge $P_{\rm mon}$ gives rise to 
a one-parameter family of JZ dyons, 
$H = \cosh(\gamma) H_{\rm mon}$,
$\Phi = \sinh(\gamma) H_{\rm mon}$, with magnetic
charge $P= \cosh(\gamma) P_{\rm mon}$ and electric charge
$Q = \cosh(\gamma) \sinh(\gamma) P_{\rm mon}$. 
This is also seen from the field equations,
\be     
\ast \Diff \ast F  =  [\Phi, \Diff \Phi] - [H, \Diff H],
\label{f-1}
\ee
\be
\ast \Diff \ast \Diff H = [\Phi,[\Phi,H]],
\label{f-2}
\ee
\be
\ast \Diff \ast \Diff \Phi = [H,[\Phi,H]],
\label{f-3}
\ee
which reduce to the monopole equations,
$\Diff \ast F = - \ast [H_{\rm mon}, \Diff H_{\rm mon}]$ 
and $\Diff \ast \Diff H_{\rm mon} = 0$, 
for $H = \cosh(\gamma) H_{\rm mon}$ and 
$\Phi = \sinh(\gamma) H_{\rm mon}$.

It is worth recalling that the total energy is not invariant 
under the transformation (\ref{boost}). However, for fixed
charges $P$ and $Q$, defined by the flux integrals
\be
P =  \int \Trace{H \, F}, \; \; \; 
Q = \int \Trace{H \ast \Diff \Phi},
\label{charges}
\ee
over the two-sphere at infinity, the energy assumes its 
global minimum for the corresponding JZ dyon solution.
This is seen as follows: Using the field equations to 
express $P$ and $Q$ as volume integrals of 
$\Trace{\Diff H \we F}$ and 
$\Trace{\Diff H \we \ast \Diff \Phi}$, respectively, the 
total energy may be expressed as follows \cite{Bogo76}, 
\cite{Coleman-77}:
\bea
E & = & \frac{1}{2} \int \left\{ (F)^{2} + (\Diff H)^{2} + 
(\Diff \Phi)^{2}+ [H, \Phi]^{2} \right\} \diff^{3}x
\nonumber\\
& = & \frac{1}{2} \int \left\{ \left(\Diff \Phi - 
\tanh (\gamma) \Diff H \right)^{2} + \left(\ast F - 
\frac{1}{\cosh^{2}(\gamma)} \Diff H \right)^{2} + 
[H, \Phi]^{2} \right\} \diff^{3}x
\nonumber\\
& + & \frac{1}{\cosh(\gamma)} \left(
Q \, \sinh(\gamma) \, + \, P \right) ,
\label{E-1}
\eea
where $\gamma$ is arbitrary and $(F)^{2}$ is a short-hand 
for $(F,F)$, etc. From this one finds the bound (assuming, 
without loss of generality, that $Q$ and $P$ are 
non-negative)
\be
E \geq \sqrt{Q^{2} + P^{2}} = \cosh^{2}(\gamma) P_{\rm mon},
\label{E-2}
\ee
where equality holds if and only if $A$, $H$ and $\Phi$ are 
subject to the first order equation 
$\Diff \Phi / \sinh(\gamma) = \Diff H / \cosh(\gamma) = 
\ast F$, which is exactly the Bogomol'nyi equation, 
\be
\ast F \, = \, \Diff H_{\rm mon} ,
\label{Bog}
\ee
written in terms of the rotated fields 
$H = \cosh(\gamma) H_{\rm mon}$ and 
$\Phi = \sinh(\gamma) H_{\rm mon}$.

\section{Linear perturbations of dyons}
\label{section-III}

The perturbation analysis for the BPS monopole is simplified
by the circumstance that the electric perturbation
$\delta \Phi_{\rm mon}$ does not couple to the magnetic 
perturbations $\delta H_{\rm mon}$ and $\delta A_{\rm mon}$.
This is an immediate consequence of the fact that the BPS 
background configuration is non-electric,
$\Phi_{\rm mon} = 0$.

Since the electric background field does not vanish for JZ 
dyons, the electric and the magnetic perturbations are 
coupled in this case. However, the linearity of the symmetry
(\ref{boost}) implies that all linear perturbations of JZ 
dyons can be obtained from the linear perturbations of the 
BPS monopole after a hyperbolic rotation with parameter 
$\sinh (\gamma) = Q/P$. It is, therefore, sufficient to 
consider the perturbation analysis of the BPS monopole. 
Before doing so, we compute the various contributions to the
angular momentum.

\subsection{Angular momentum}
\label{subsection-III-A}

The total angular momentum (along the symmetry axis)
of a stationary YMH configuration is
\be
J = \int T_{t\varphi}\, \diff^3 x ,
\label{J-1}
\ee
where the relevant component of the stress-energy tensor 
in terms of the three-dimensional quantities is given by
\bdm
T_{t\varphi}= \frac{1}{2} \Trace{[\Phi,H] \Diff H - 
\ast (\Diff \Phi \we \ast F)}_{\varphi} .
\edm
By virtue of the field equation (\ref{f-1}) and the 
relations $\Trace{\Phi [\Phi, \Diff \Phi]} = 0$ and
$\Trace{\Phi [H, \Diff H]}$ $=$ $\Trace{[\Phi, H] \Diff H]}$,
we also find (after integrating by parts)
\be
T_{t\varphi}= - \frac{1}{2} \left( 
\ast \diff \, \Trace{\Phi \ast F} \right)_{\varphi} .
\label{J-2}
\ee
This shows that both the electric and the magnetic 
perturbations of JZ dyons contribute to the angular 
momentum, since
\bdm
\delta T_{t\varphi}= - \frac{1}{2} \left( \ast \diff \, 
\Trace{\delta \Phi \ast F + \Phi \ast \delta F} 
\right)_{\varphi} .
\edm
(Note that the second term is absent if the electric
background field vanishes, implying that only electric 
perturbations give rise to the angular momentum of a BPS 
monopole.) Since the dyon perturbations can be obtained from
the monopole perturbations, we express the angular momentum
in terms of the latter, using
$\Phi = \sinh(\gamma) H_{\rm mon}$ and
$\delta \Phi = \sinh(\gamma) \delta H_{\rm mon} + 
\cosh(\gamma) \delta \Phi_{\rm mon}$.
With
\be
\delta T_{t\varphi} \, = \,  \cosh(\gamma) \, 
\delta T^{\rm el}_{t\varphi} + \sinh(\gamma) \,
\delta T^{\rm mg}_{t\varphi},
\label{T-el-mg}
\ee
one finds
\bea
\delta T^{\rm el}_{t\varphi} & = & - \frac{1}{2}
\left( \ast \diff \Trace{\delta \Phi_{\rm mon} \ast F} 
\right)_{\varphi} ,
\label{T-el} \\
\delta T^{\rm mg}_{t\varphi} & = & - \frac{1}{2}
\left( \delta \ast \diff \Trace{ H_{\rm mon} \ast F} 
\right)_{\varphi} .
\label{T-mg}
\eea
It is worthwhile noticing that both contributions to 
$\delta T_{t\varphi}$ are separately gauge invariant. This
is obvious for the electric part, since this is proportional
to the perturbation of a field which vanishes on the 
background, namely $\Phi_{\rm mon}$. The same is true for 
the magnetic part, since the quantity 
$\diff \Trace{H_{\rm mon} \ast F}$ vanishes as well for
a PBS background configuration. (Use 
$\ast F = \Diff H_{\rm mon}$ to see this.) In fact, defining
the one-form $B$ according to
\be
B \, = \, \Diff H_{\rm mon} - \ast F,
\label{def-B}
\ee
the magnetic contribution (\ref{T-mg}) 
to the angular momentum can be cast into the simple form
\be
\delta T^{\rm mg}_{t\varphi} = - \frac{1}{2}
\left( \ast \Trace{ \delta B \we \ast F} \right)_{\varphi} ,
\label{T-mg-2}
\ee
which is manifestly gauge invariant, since, by definition,
$B$ vanishes for the BPS background configuration.

The above expressions imply the following facts: First, the
perturbation analysis for JZ dyons reduces to the perturbation
analysis for BPS monopoles. Second, the electric {\it and\/}
the magnetic perturbations of a BPS background contribute to
the dyon angular momentum. Third, only the 
{\it non self-dual\/} modes, that is, the magnetic 
perturbations with $\delta B \neq 0$ contribute to the dyon 
angular momentum. 

The last statement reveals a fundamental difference between 
the perturbation theory of BPS monopoles and JZ dyons:
Although the perturbation equations for JZ dyons can be 
reduced to the ones for the BPS monopole, the physical 
contents is quite different: While only electric 
perturbations can give rise to the angular momentum of a 
monopole configuration, magnetic perturbations need to be
taken into account as well in the dyon case. Moreover, it is
not sufficient to consider perturbations within the 
Bogomol'nyi sector, since the latter cannot contribute to 
the angular momentum of a dyon.

\subsection{Linear perturbations of the BPS monopole}
\label{subsection-III-B}

Since the perturbation analysis of the JZ dyons can be 
reduced to the one for the BPS monopole, we shall now focus 
on the latter. In the following we omit the 
subscript ``${\rm mon}$'' indicating the monopole fields, that 
is, we write $\delta H$ for $\delta H_{\rm mon}$, etc.  
Suppose that there is (at least) a one-parameter family of
continuous deformations of the BPS monopole background, 
$\ast F = \Diff H$, $\Phi = 0$. Then the tangent to this
satisfies the linearized field equations. In order to 
linearize Eqs. (\ref{f-1}) and (\ref{f-2}), it is very 
convenient to introduce the one-form field $B$ defined in 
Eq. (\ref{def-B}). One may then write the first field 
equation in the form $\Diff B = \Diff^{2} H - \Diff \ast F =
[H, \ast \Diff H - F] - [\Phi,\ast \Diff \Phi]$, whereas 
the second field equation becomes $\Diff \ast B = 
\Diff \ast \Diff H - \Diff F = \ast [\Phi,[\Phi,H]]$. Hence,
Eqs. (\ref{f-1}) and (\ref{f-2}) assume the form
\be
\Diff B - [H  , \ast B] = - [\Phi,\ast \Diff \Phi] ,
\label{B-1}
\ee
and
\be
\Diff \ast B = \ast [\Phi,[\Phi,H]] ,
\label{B-2}
\ee
respectively.
The linearization of the field equations (\ref{f-3}),
(\ref{B-1}), (\ref{B-2}) is completely trivial, since both 
the electric field $\Phi$ and the magnetic one-form 
$B \equiv \Diff H - \ast F$ vanish for a BPS background. Hence, 
the linearized field equations involve only the gauge 
invariant perturbations $\delta \Phi$ and $\delta B$: One 
immediately finds the result
\bea
\mbox{electric perturbations:} & &
\; \; \Diff \ast \Diff \delta \Phi = \ast 
[H,[\delta \Phi,H]] ,
\label{deltaphi}\\
\mbox{magnetic perturbations:} & &
\; \; \Diff \delta B = [H , \ast \delta B] , \; \; \; 
\Diff \ast \delta B = 0 ,
\label{deltab} 
\eea
where $\delta B$ is obtained from the definition 
(\ref{def-B}), that is
\be                                       
\delta B = \Diff \delta H -\ast \Diff \delta A - 
[H,\delta A] .
\label{Inh-Bog}
\ee
(Here and in the following all quantities without 
a ``$\delta$'' refer to background fields.) 
Before we consider the harmonic analysis of Eqs. 
(\ref{deltaphi})--(\ref{Inh-Bog}), we note the following:
\begin{itemize}
\item
The linearization of the Bogomol'nyi equation (\ref{Bog}),
$\delta B = 0$, has been studied extensively in the 
literature. The solutions to $\delta B = 0$ are, however,
only a subset of the general magnetic perturbations. The 
full magnetic perturbations are governed by the second order
equations for $\delta A$ and $\delta H$, which are 
equivalent to the first oder equations (\ref{deltab}) for 
$\delta B$ and the inhomogeneous equation (\ref{Inh-Bog}).
In particular, we have already argued above that only the 
non-trivial solutions $\delta B \neq 0$ to 
Eq. (\ref{deltab}) can contribute to the angular momentum 
[see Eq. (\ref{T-mg-2})].

\item
In order to find the general magnetic perturbations, one 
proceeds in two steps: First, one has to solve the system 
(\ref{deltab}) for $\delta B$. Once $\delta B$ is known, it 
remains to solve the inhomogeneous linearized Bogomol'nyi 
equation (\ref{Inh-Bog}) for $\delta A$ and $\delta H$. This
is achieved by using Green's method, also taking advantage 
of the explicitly known solutions to the homogeneous 
equations, $\delta B = 0$, derived in \cite{Adler79} and 
\cite{Akhoury80}.

\item
Since the background BPS configuration has vanishing $B$, 
the magnetic perturbation $\delta B$ is manifestly gauge 
invariant. This is also verified by using the general 
behavior of the perturbations $\delta A$ and $\delta H$ 
under gauge transformations generated by a Lie algebra 
valued scalar field $\chi$:
\be
\delta A\rightarrow\delta A+ \Diff\chi,\ \ \ \ \
\delta H\rightarrow\delta H+[H,\chi] .
\label{16}
\ee
Hence, $\delta \Diff H \rightarrow \delta \Diff H  + 
[\Diff H, \chi]$, and $\delta F \rightarrow \delta F + 
[F,\chi]$, implying that 
$\delta B \rightarrow \delta B + [B,\chi] = \delta B$.

\item
The second equation in (\ref{deltab}) is a consistency condition for 
the first one: Indeed, applying $\Diff$ on the first equation and
using $\Diff^{2} \delta B = [F, \delta B]$ on the LHS, and 
$[\Diff H, \ast \delta B] = [\ast \Diff H, \delta B] = 
[F, \delta B]$ on the RHS, yields the necessary condition
$[H,\Diff \ast \delta B]=0$.
\end{itemize}

\section{Harmonic analysis}
\label{section-IV}

Since the unperturbed BPS solution is spherically symmetric,
we perform a multipole decomposition and rewrite the 
electric perturbation equations (\ref{deltaphi}) and the 
magnetic ones (\ref{deltab}), (\ref{Inh-Bog}) as systems of 
ordinary differential equations with respect to the radial 
coordinate. Using these equations, we show that the angular 
momentum integral can be computed exactly. Hence, the total 
angular momentum arising from electric and magnetic 
perturbations is determined by the asymptotic behavior of 
the gauge invariant amplitudes $\delta \Phi$ and $\delta B$,
respectively.

\subsection{Isospin harmonics}
\label{subsection-IV-A}

The basic fields $H$, $\Phi$, $A$, the auxiliary field
$B$, and their perturbations, are functions and one-forms
with values in the Lie algebra su(2) of the gauge
group SU(2). Let us start by considering such functions on
the two-sphere $S^{2}$. A convenient basis,
reducing the natural representation of SU(2), is obtained by
taking the inner product of the vector spherical harmonics 
$\YLJM$ with the basis $\taubold = \sigbold/(2i)$ of su(2) 
(where $\sigbold$ are the Pauli matrices) 
\be
\CLJM(\vartheta,\varphi) = 
\taubold \cdot \YLJM(\vartheta,\varphi) .
\label{HA-1}
\ee
The isospin harmonics $\CLJM$ have total angular momentum 
$J$ and fixed parity $(-1)^{L}$. Instead of the 
$\YLJM$ it is also usual to consider the basis 
$\Yarg{(\lambda)}$ (with $\lambda = 0, \pm 1$).
For $\lambda = 0$ and $\lambda = 1$
these vector harmonics are transverse, 
while they are longitudinal for $\lambda = -1$ (with respect
to the radial unit direction $\hat{\radbold}$). The 
transverse harmonics $\Yarg{(1)}$ and $\Yarg{(0)}$ are also 
called electric and magnetic multipoles, respectively. They 
are obtained by applying certain differential 
operators on the ordinary spherical harmonics $Y_{JM}$, 
while the longitudinal harmonics are given by 
$\Yarg{(-1)} = \hat{\radbold} Y_{JM}$ 
(see, e.g., \cite{Varshalovich} or \cite{Edmonds}). The
formulas for the $\Yarg{(\lambda)}$ can readily be 
translated into the corresponding formulas for
the isospin harmonics 
$\Carg{(\lambda)} = \taubold \cdot \Yarg{(\lambda)}$
(with $\lambda = 0, \pm 1$). One finds
\bea
\Carg{(-1)} & = & \taur Y_{JM} ,
\nonumber\\
\Carg{(0)} & = & \frac{i}{\sqrt{J(J+1)}}
\sprod{\diff \taur}{\hatast \diff Y_{JM}} ,
\label{HA-2} \\
\Carg{(+1)} & = & \frac{1}{\sqrt{J(J+1)}}
\sprod{\diff \taur}{\diff Y_{JM}} ,
\nonumber
\eea
where $\taur = \taubold \cdot \hat{\radbold}$.
Here $\sprod{\,}{\,}$ and $\hatast$ denote the inner product
and the Hodge dual with respect to the standard metric on $S^{2}$. 
(Also note that the spherical components of the $\taubold$
obey the relations 
$\diff\taur = \tautheta \diff\vartheta + \tauphi \sin \! \vartheta \diff\varphi$
and $[\tautheta, \tauphi] = \taur$; see Appendix 
\ref{AA-1}.) In terms of the isospin harmonics, the 
well-known relations between the vector harmonics 
$\YLJM$ and $\Yarg{(\lambda)}$ become
\bea
\Carg{J+1} & = & \frac{1}{\sqrt{2J+1}} \left[
\sqrt{J} \Carg{(1)} - \sqrt{J+1} \Carg{(-1)} \right],
\nonumber\\
\Carg{J}   & = & \Carg{(0)} ,
\label{HA-3} \\
\Carg{J-1} & = & \frac{1}{\sqrt{2J+1}} \left[
\sqrt{J+1} \Carg{(1)} + \sqrt{J} \Carg{(-1)} \right].
\nonumber
\eea
By construction, the isospin harmonics $\Carg{J}$, $\Carg{J \pm 1}$
are eigenfunctions of the spherical
Laplacian, $\hatlap = \hatast\diff\hatast\diff$, and of the parity
operator, $\hat{P}$:
\be
\hatlap \, \CLJM = - L (L+1) \,  \CLJM ,
\label{Laplacian}
\ee
\be
\hat{P} \, \CLJM = (-1)^{L} \, \CLJM , 
\label{Parity}
\ee
where $L = J, J \pm 1$.
(The exterior derivatives of the isospin harmonics
$\Carg{J}$ and $\Carg{J \pm 1}$ and their $S^{2}$ duals
are particularly convenient for analyzing perturbations of 
Lie algebra valued one-forms, \cite{OB}, \cite{BH97}. For 
the general theory of monopole harmonics we refer to 
\cite{Weinberg93}.)

\subsection{Perturbation Amplitudes}
\label{subsection-IV-B}

Since rotational modes are our primary concern in this 
paper, we now focus on the sector $J=1$. For the 
$C^{(\lambda)}_{1\,0}$ ($\lambda = 0, \pm 1$) we use 
(with some change of normalization)
the letters $X$, $Y$ and $Z$. A convenient basis of $J=1$ 
isospin harmonics then is
\bea
X & = & \taur \, K, \; \; 
\mbox{where $\; K \equiv \cos \! \vartheta$},
\nonumber\\
\sqrt{2} \, Y & = & \sprod{\diff\taur}{\diff K} = -\tautheta 
\sin \! \vartheta,
\nonumber\\
\sqrt{2} \, Z & = & - \sprod{\diff\taur}{\hatast \diff K} = \tauphi 
\sin \! \vartheta,
\label{XYZ}
\eea
where  $X$ and $Y$ span the even parity sector, while $Z$ 
has odd parity. The su(2) valued electric perturbation 
function $\delta \Phi$ can, therefore, be expanded
as $\delta \Phi = \delta \Phi^{\rm even} + \delta \Phi^{\rm odd}$, with
\bea
\delta \Phi^{\rm even} & = & \frac{1}{r} \left( \phi_{-} X + 
\phi_{+} Y \right) ,
\nonumber\\
\delta \Phi^{\rm odd} & = & \frac{1}{r} \left( \tilde{\phi} Z \right) .
\label{exp-Phi}
\eea
(The factor $1/r$ is introduced for convenience; 
see, e.g. Eqs. (\ref{el-even}), (\ref{el-odd}).
Throughout this paper, all amplitudes furnished with a 
tilde refer to the odd parity sector, which is relevant
for deformations.)
A similar expansion holds for $\delta H$; however, unlike
$\delta \Phi$, $\delta H$ is not gauge invariant; see
Sect. \ref{subsection-V-C} and Appendix \ref{AA-4}.

Turning to Lie algebra valued one-forms, we note that the 
exterior derivatives of the basis functions $X$, $Y$ and $Z$
can be expressed in terms of the derivatives of $\taur$ and 
$K = \cos \! \vartheta \propto Y_{1\,0}$. (This is a 
peculiarity of the $J=1$ harmonics, for which 
$\diff C^{(0)}_{1\,M} = (\sqrt{2} \diff Y + 
\diff X)/\sqrt{3} = 0$.)
One finds
\bea
\diff X = -\sqrt{2} \, \diff Y 
& = & \taur \, \diff K + K \, \diff \taur ,
\nonumber\\
\hatast \sqrt{2} \, \diff Z 
& = & \taur \, \diff K - K \, \diff \taur .
\label{dXdYdZ}
\eea
As the parity operation commutes with the exterior 
differentiations and anti-commutes with the Hodge dual, one 
can expand the
su(2) valued magnetic perturbation one-form $\delta B$
as $\delta B = \delta B^{\rm even} + \delta B^{\rm odd}$, 
with
\bea
\delta B^{\rm even} & = & \frac{1}{r^{2}} \left( b_{-} X + 
b_{+} Y \right)
\diff r + \beta_1 \, \taur \diff K + \beta_2 \, K 
\diff \taur ,
\nonumber\\
\delta B^{\rm odd} & = & \frac{1}{r^{2}} ( \tilde{b} Z )
\diff r + \tilde{\beta}_1 \, \hatast \taur \diff K + 
\tilde{\beta}_2 \, \hatast K \diff \taur ,
\label{exp-B}
\eea
where $\tilde{b}$, $b_{\pm}$, $\beta_{1,2}$ and
$\tilde{\beta}_{1,2}$ depend on the radial coordinate $r$.
(Again, a similar formula holds for $\delta A$. In contrast
to $\delta B$, $\delta A$ is not gauge invariant, implying 
that not all coefficients in the expansion of $\delta A$ 
correspond to physical degrees of freedom; see 
Sect. \ref{subsection-V-C} and Appendix \ref{AA-4}).

At this point we also recall that the background
gauge potential and Higgs field are parametrized 
in terms of two radial functions $w(r)$ and $h(r)$
(see Appendix \ref{AA-1}),
\be
A = \left[ 1 - w(r) \right] \hatast \diff \taur , \; \; \; 
H = h(r) \, \taur .
\label{A-H}
\ee
Since $\taur$ is an eigenfunction of the spherical 
Laplacian,
$\diff \hatast \diff \taur = -2 \taur \diff \Omega$,
the background field strength becomes
$F = -\diff w \we \hatast \diff \taur + 
(w^{2}-1) \taur \diff \Omega$. The BPS equations,
$F = \ast \Diff H$, thus read
\be
w' = w\, h , \; \; \; r^{2} h' = w^2 - 1 ,
\label{wh-eq}
\ee
with the globally regular solution
\be
w(r) = \frac{r}{\sinh(r)} , \; \; \; h(r) = 
\frac{1}{r} - 
\frac{\cosh(r)}{\sinh(r)} .
\label{w-h}
\ee
For later use we also note that the second order equation
for $h$ can be integrated, which yields the useful relation
\be
h' =  h^2 - \frac{2 h}{r} - 1 .
\label{wh-eq-2}
\ee

\subsection{Integration of angular momentum}
\label{subsection-IV-C}

We now show that the total angular momentum 
$\delta J = \cosh (\gamma) \delta J^{\rm el} + 
\sinh (\gamma) \delta J^{\rm mg}$ can be expressed in terms 
of the values of the gauge invariant perturbations
$\delta \Phi$ and $\delta B$ at the origin and at infinity.
According to Eqs. (\ref{T-el}) and (\ref{T-mg-2}), 
the electric and the magnetic perturbations give rise to
\be
\delta J^{\rm el} = -\frac{1}{2} \int
\left( \ast \diff \Trace{\delta \Phi^{\rm even} \ast F} 
\right)_{\varphi} \diff^{3}x
\label{J-el}
\ee
and
\be
\delta J^{\rm mg} = -\frac{1}{2} \int
\left( \ast \Trace{ \delta B^{\rm even} \we \ast F} 
\right)_{\varphi} \diff^{3}x \, ,
\label{J-mg}
\ee
respectively.
Here we have already used the fact that only the even parity
sector contributes to the total angular momentum.
In order to express the above integrands in terms of the
radial amplitudes $\phi_{\pm}$, $b_{\pm}$ and $\beta_{1,2}$,
we first note that the background field strength can be 
written in the simple form
\be
\ast F = w' \diff \taur + h' \taur \diff r \, .
\label{FBG}
\ee
Taking advantage of the trace formulas
$\Trace{X \taur} = -K/2$,
$\Trace{Y \taur} = \Trace{Z \taur} = 0$, and
$\Trace{X \diff \taur} = 0$,
$\Trace{Y \diff \taur} = \hatast \Trace{Z \diff \taur} 
= -dK/(2\sqrt{2})$, 
it is now not difficult to compute the above integrands from
the expansions (\ref{exp-Phi}) and (\ref{exp-B}). One finds
\bea
\ast \diff \Trace{\delta \Phi^{\rm even} \ast F} & = &
\frac{1}{2 r^{2}} \left[
r h' \phi_{-} - \frac{r^{2}}{\sqrt{2}} \left(
\frac{w' \phi_{+}}{r} \right)' \, \right] \hatast \diff K ,
\label{br-electric}\\
\ast \Trace{\delta B^{\rm even} \we \ast F} & = &
\frac{1}{2 r^{2}} \left[
r^{2} h' \beta_{1} - \frac{1}{\sqrt{2}} w' b_{+}
\right] \hatast \diff K .
\label{br-magnetic}
\eea
With $K \equiv \cos\!\vartheta$ we have
$\hatast \diff K = - \sin^{2}\!\vartheta \diff \varphi$, 
which shows that the above formulas yield the 
$\varphi$ components
appearing in the integrands of Eqs. (\ref{J-el})
and (\ref{J-mg}). It is an interesting fact that the 
above brackets can be written as radial derivatives. 
This enables one
to perform the angular momentum integrals and to express
$\delta J^{\rm el}$ and $\delta J^{\rm mg}$ in terms of
the values of $\delta \Phi^{\rm even}$ and 
$\delta B^{\rm even}$
at the origin and at infinity. In order to see this, one
has to use the perturbation equations in the harmonic
decomposition, as given in the next section. Considering the
magnetic part, one uses the first two equations in 
(\ref{even-B-eqs}) to obtain 
$2(1-w^{2}) \beta_{1}=b'_{-}+\sqrt{2}w b'_{+}$, 
which enables one to eliminate $\beta_{1}$ in 
Eq. (\ref{br-magnetic}). 
Also taking advantage of the background equation 
(\ref{wh-eq}), one then has 
$[r^{2} h' \beta_{1} - w' b_{+} / \sqrt{2}] = 
-[b_{-}+ \sqrt{2} w b_{+}]'/2$. A similar, but more
complicated manipulation uses the second order equations
(\ref{el-even}) to write the electric contribution 
(\ref{br-electric}) in the desired form; 
see Appendix \ref{AA-6}. 
The two contributions (\ref{J-el}) and (\ref{J-mg}) to the 
angular momentum finally become
\bea
\delta J^{\rm el}  & = &  -\frac{\pi}{3} 
\left[
\left(1-w^{2}-2rh \right) \phi_{-} + r^{2} h \phi'_{-} +
\sqrt{2} w r h \phi_{+} \right]^{\infty}_{0} , 
\label{JEL}\\
\delta J^{\rm mg}  & = &  - \frac{\pi}{3} 
\left[ b_{-} + \sqrt{2} w b_{+} \right]^{\infty}_{0} .
\label{JMG}
\eea

\section{Perturbation equations}
\label{section-V}

Using the expansions (\ref{exp-Phi}) and (\ref{exp-B}), as 
well as the tools developed in Appendix \ref{AA-2} and 
\ref{AA-3}, it is now straightforward
to write down the system of perturbation equations. 
This consists of Eq. (\ref{deltaphi}) for the electric 
perturbations
$\delta \Phi$, Eqs. (\ref{deltab}) for the magnetic 
perturbations $\delta B$, and the inhomogeneous BPS Eqs. 
(\ref{Inh-Bog}) for $\delta H$ and $\delta A$.

\subsection{Electric perturbations}
\label{subsection-V-A}

For the electric perturbations (\ref{exp-Phi}), 
governed by Eq. (\ref{deltaphi}), one finds the 
differential equations
\be
\left( \begin{array}{c}
\phi''_{-} \\ \phi''_{+} \end{array} \right) = 
\frac{1}{r^2} \left( \begin{array}{cc}
2 (w^{2}+1) & -2 \sqrt{2} w \\
-2 \sqrt{2} w & (w^{2}+1+r^2 h^2)
\end{array} \right) \,
\left( \begin{array}{c}
\phi_{-} \\ \phi_{+} \end{array} \right)
\label{el-even}
\ee
in the even parity sector, and
\be
\tilde{\phi}'' \, = \,  
\frac{1}{r^2} \left(w^{2} + 1 + r^2 h^{2} \right) 
\tilde{\phi}
\label{el-odd}
\ee
in the odd parity sector. Here we have used Eq. (\ref{A13})
to compute the LHS of Eq. (\ref{deltaphi}), and 
$[\taur,X] = 0$, $[\taur,Y] = -Z$, $[\taur,Z] = Y$
to obtain the RHS: 
$[H,[\delta \Phi, H]] = h^{2} r^{-1} (\phi_{+} Y + 
\tilde{\phi} Z)$.

\subsection{Magnetic perturbations: $\delta B$ equations}
\label{subsection-V-B}

In order to determine the magnetic perturbations $\delta B$, 
we first write the decomposition (\ref{exp-B}) in the form
\be
\delta B = \frac{1}{r^2} \, b \, \diff r + \hat{B} ,
\label{BbB}
\ee
where the one-form $\hat{B}$ is tangential to $S^2$.
In terms of $b$ and $\hat{B}$, the magnetic perturbation
equations (\ref{deltab}) assume the form
\bdm
[H,b] = \hatast \hatDiff \hat{B} ,
\edm
\bdm
\hat{B}' = r^{-2} \hatDiff b -  [H,\hatast \hat{B}] ,
\edm
\be
b' = - \hatast \hatDiff \hatast \hat{B}.
\label{BbB-eqs}
\ee
Here we have used the fact that the unperturbed
gauge potential has no radial component, 
implying the decomposition 
$\Diff = \diff r \we \partial_r + \hatDiff$ for the 
covariant derivative (see Appendix \ref{AA-2}).
Taking advantage of the formulas given in
Appendix \ref{AA-3}, it is now 
not hard to obtain the sets of differential 
equations for the radial functions parametrizing 
$\delta B^{\rm even}$ and $\delta B^{\rm odd}$.
One finds
\bdm
b'_{-} = 2 (\beta_1 + w \beta_2) , \; \; \; 
b'_{+} = -\sqrt{2} (w \beta_{1} +  \beta_{2}) ,
\edm
\bdm
r^2 \beta_1' = b_{-} - 
\frac{w}{\sqrt{2}} \, b_{+} , \; \; \; 
r^2 (\beta_2' + h \beta_2) = w b_{-} - 
\frac{1}{\sqrt{2}} \, b_{+} , 
\edm
\be
h \, b_{+} = \sqrt{2} (\beta_2 - w \beta_1) 
\label{even-B-eqs}
\ee
for the even parity sector, and
\bdm
\tilde{b}' = \sqrt{2} (w \tilde{\beta}_1 -  
\tilde{\beta}_2) ,
\edm
\bdm
r^2 \tilde{\beta}_1' = \frac{w}{\sqrt{2}} \, \tilde{b} , 
\; \; \; 
r^2 (\tilde{\beta}_2' + h \tilde{\beta}_2) = - 
\frac{1}{\sqrt{2}} \, \tilde{b} ,
\edm
\be
0 = \tilde{\beta}_1 + w \tilde{\beta}_2 , \; \; \;  
h \, \tilde{b} = \sqrt{2} (w \tilde{\beta}_1 + 
\tilde{\beta}_2)
\label{odd-B-eqs}
\ee
for the odd parity sector. We note that both sectors
contain constraint equations, reflecting the fact that 
the second equation in (\ref{deltab}) is an integrability
condition for the first one.

At this point we also note the following, somewhat 
surprising fact:
The scalar magnetic amplitudes $b_{\pm}$ and $\tilde{b}$ 
satisfy the same set of second order equations 
(\ref{el-even}), (\ref{el-odd}) 
as the electric amplitudes $\phi_{\pm}$ and $\tilde{\phi}$,
\be
\left( \begin{array}{c}
b''_{-} \\ b''_{+} \end{array} \right) = 
\frac{1}{r^2} \left( \begin{array}{cc}
2 (w^{2}+1) & -2 \sqrt{2} w \\
-2 \sqrt{2} w & (w^{2}+1+r^2 h^2)
\end{array} \right) \,
\left( \begin{array}{c}
b_{-} \\ b_{+} \end{array} \right) ,
\label{b-even}
\ee
\be
\tilde{b}'' \, = \,  
\frac{1}{r^2} \left(w^{2} + 1 + r^2 h^{2} \right) \tilde{b} .
\label{b-odd}
\ee
This follows from the arguments given in 
Appendix \ref{AA-2}, and is also verified directly from the 
above equations. Using the
odd parity equations (\ref{odd-B-eqs}) we have, for 
instance $\tilde{b}'' = \sqrt{2}(w' \tilde{\beta}_{1} + 
w \tilde{\beta}'_{1} - \tilde{\beta}'_{2})$
$=$ $[h^2 + r^{-2}(w^2 + 1)] \tilde{b}$, where we have 
used the Bogomol'nyi equations (\ref{wh-eq}) 
for the background fields $h$ and $w$.

We also point out that not all solutions to the second
order equations (\ref{b-even}) and (\ref{b-odd}) satisfy the
first order equations (\ref{even-B-eqs}) and 
(\ref{odd-B-eqs}). In fact, 
it is not hard to see that the solution spaces 
defined by Eqs. (\ref{even-B-eqs}) and (\ref{odd-B-eqs})
are three- and one-dimensional, respectively, rather than
four- and two-dimensional.
 
\subsection{Magnetic perturbations: Inhomogeneous 
BPS equations}
\label{subsection-V-C}

In order to write out the inhomogeneous Bogomol'nyi 
equations (\ref{Inh-Bog}), we need the harmonic 
decomposition of the fields $\delta H$ and $\delta A$.
The fact that the latter are not gauge invariant enables us 
to get rid of certain amplitudes. In Appendix \ref{AA-4} 
it is shown that -- up to a pure gauge -- the harmonic 
decompositions of $\delta H$ and $\delta A$ assume the form
\bea
\delta H^{\rm even} & = & \gamma_{-} X + \gamma_{+} Y ,
\nonumber \\
\delta H^{\rm odd} & = & \tilde{\gamma} Z,
\label{delta-H}
\eea
and
\bea
\delta A^{\rm even} & = & 
\alpha_1 \hatast \taur \diff K + 
\alpha_2 \hatast K \diff \taur ,
\nonumber\\
\delta A^{\rm odd} & = & 
\tilde{\alpha}_1 \, \taur \diff K + 
\tilde{\alpha}_2  K \diff \taur ,
\label{delta-A}
\eea
respectively. 
The radial functions $\gamma_{\pm}$, $\tilde{\gamma}$, 
$\alpha_{1, 2}$ and $\tilde{\alpha}_{1,2}$ 
are gauge invariant, up to a one-dimensional set of residual 
gauge transformations in the even parity sector,
\bea
\gamma_{-} & \rightarrow & \gamma_{-} , \; \; \; 
\gamma_{+} \rightarrow \gamma_{+} + h \, c_{3} ,
\nonumber\\
\alpha_{1} & \rightarrow & \alpha_{1} + 
\frac{w}{\sqrt{2}} \, c_{3} ,
\nonumber\\
\alpha_{2} & \rightarrow & \alpha_{2} - 
\frac{1}{\sqrt{2}}  c_{3}, 
\label{res-even}
\eea
and a two-dimensional set of residual gauge transformations
in the odd parity sector,
\bea
\tilde{\gamma} & \rightarrow & \tilde{\gamma} - h \, c_{2} ,
\nonumber\\
\tilde{\alpha}_{1} & \rightarrow & \tilde{\alpha}_{1} + 
c_{1} - \frac{w}{\sqrt{2}} \, c_{2} ,
\nonumber\\
\tilde{\alpha}_{2} & \rightarrow & \tilde{\alpha}_{2} + 
w \, c_{1} - \frac{1}{\sqrt{2}} \, c_{2} ,
\label{res-odd}
\eea
where $c_{1}$, $c_{2}$ and $c_{3}$ are arbitrary constants
parametrizing the residual gauge freedom 
(see Appendix \ref{AA-4}).
In terms of the gauge invariant source terms $b_{\pm}$, 
$\tilde{b}$, $\beta_{1, 2}$ and $\tilde{\beta}_{1,2}$, and 
the (almost) gauge invariant amplitudes introduced above, 
the inhomogeneous linearized
Bogomol'nyi equations (\ref{Inh-Bog}) eventually become
\bea
r^{2} \gamma'_{-} + 2 (\alpha_{1} + w \alpha_{2}) 
& = & b_{-} , 
\nonumber\\
r^{2} \gamma'_{+} / \sqrt{2} - (w \alpha_{1} + \alpha_{2}) 
& = & b_{+} / \sqrt{2} , 
\nonumber\\
\alpha'_{1} + \gamma_{-} - w \gamma_{+} / \sqrt{2} 
& = & \beta_{1} ,
\nonumber\\
\alpha'_{2} - h \alpha_{2} + w \gamma_{-} - 
\gamma_{+} / \sqrt{2} & = & \beta_{2} ,
\label{Inh-Bog-even}
\eea
in the even parity sector, and
\bea
r^{2} \tilde{\gamma}' / \sqrt{2} + 
(\tilde{\alpha}_{2} - w \tilde{\alpha}_{1})
& = & \tilde{b} /\sqrt{2} ,
\nonumber\\
- \tilde{\alpha}'_{1} + w \tilde{\gamma} / \sqrt{2}
& = & \tilde{\beta}_{1} ,
\nonumber\\
- \tilde{\alpha}'_{2}  + h \tilde{\alpha}_{2} - \tilde{\gamma} / \sqrt{2}
& = & \tilde{\beta}_{2} ,
\label{Inh-Bog-odd}
\eea
in the sector with odd parity.

For vanishing RHS, the above equations are the 
linearized Bogomol'nyi
equations, which have been studied in the literature.
Using the background equations (\ref{wh-eq}), it is easy 
to verify that the residual gauge mode
$\gamma_{-} = 0$, $\gamma_{+} = 
\sqrt{2} h$, $\alpha_{1} = w$, $\alpha_{2} = -1$ 
satisfies the homogeneous equations (\ref{Inh-Bog-even}),
while the residual gauge modes
$\tilde{\gamma} = \sqrt{2}h$, $\tilde{\alpha}_{1} = w$, 
$\tilde{\alpha}_{2} = 1$ and
$\tilde{\gamma}= 0$, $\tilde{\alpha}_{1} = 1$, 
$\tilde{\alpha}_{2} = w$ are solutions to
the homogeneous equations (\ref{Inh-Bog-odd}).

\section{Odd parity modes}
\label{section-VI}

We shall now solve the perturbation equations. We start 
with the odd parity sector, for which all solutions can be 
obtained in closed form. We emphasize, however, that this 
sector is of minor importance, since the odd parity modes 
cannot contribute to the angular momentum. 
In Sect. \ref{subsection-VI-A} we compute the magnetic amplitudes 
$\delta B^{\rm odd}$, which we use in
Sect. \ref{subsection-VI-B} as source terms to obtain the 
perturbations $\delta H^{\rm odd}$ and $\delta A^{\rm odd}$. 
In Sect. \ref{subsection-VI-C} we finally compute the 
electric perturbations $\delta \Phi^{\rm odd}$.

\subsection{Solutions to the $\delta B$ equations}
\label{subsection-VI-A}

In order to compute the source term $\delta B^{\rm odd}$, 
we have to solve Eqs. (\ref{odd-B-eqs}) for the amplitudes 
$\tilde{b}$ and $\tilde{\beta}_{1,2}$ defined in 
Eq. (\ref{exp-B}). Using the last two equations in
(\ref{odd-B-eqs}) to express $\tilde{\beta}_{1}$ and 
$\tilde{\beta}_{2}$ in terms of $\tilde{b}$, the first 
equation becomes 
$\tilde{b}'/\tilde{b} = h (w + w^{-1})/(w-w^{-1})$,
which is trivial to solve, since the numerator is the 
derivative of the denominator. Hence, the only solution to 
Eqs. (\ref{odd-B-eqs}) is
\be
\tilde{b} = w - \frac{1}{w}, \; \; \; 
\sqrt{2} \tilde{\beta}_1 = h , \; \; \; 
\sqrt{2} \tilde{\beta}_2 = - \frac{h}{w} \, .
\label{sol-odd-B-eqs}
\ee
Inserting this back into the expansion (\ref{exp-B}), and 
using the background equation (\ref{wh-eq}) for $h'$ and 
the formula (\ref{A13}) for $\Diff Z$, yields the simple 
result
\be
\delta B^{\rm odd} = \frac{1}{w} \, \Diff (h \, Z) \, .
\label{sol-odd-B-eqs-bis}
\ee

\subsection{Solutions to the inhomogeneous BPS equations}
\label{subsection-VI-B}

Now that the source terms for the linearized 
Bogomol'nyi equations (\ref{Inh-Bog-odd}) are known, 
we can proceed and solve the inhomogeneous problem. 
Since the homogeneous equations admit three solutions, 
two of which are the residual gauge modes 
$\tilde{\gamma} = \sqrt{2}h$, $\tilde{\alpha}_{1} = w$, 
$\tilde{\alpha}_{2} = 1$ and
$\tilde{\gamma}= 0$, $\tilde{\alpha}_{1} = 1$, 
$\tilde{\alpha}_{2} = w$,
we need to find the remaining solution of the
homogeneous problem and a solution of the 
inhomogeneous equations. This is achieved by 
deriving a third order equation for $\tilde{\gamma}$. 
In fact, since $\tilde{\gamma}= 0$
is a residual gauge mode of Eqs. (\ref{Inh-Bog-odd}), the 
differential equation for
$\tilde{\gamma}$ will be of second, rather than third order.
Moreover, using the second residual gauge mode, one 
eventually ends up with a first order equation. First,
one easily finds from Eqs. (\ref{Inh-Bog-odd})
\be
\left(r^{2} \tilde{\gamma}'\right)' - 
h \left(r^{2} \tilde{\gamma}'\right) - 
(w^{2} + 1) \tilde{\gamma} = -h \, \tilde{b} ,
\label{gamma-1}
\ee
where $\sqrt{2}(w \tilde{\beta}_1 - \tilde{\beta}_2) = \tilde{b}'$
was used on the RHS. Now using the second residual gauge mode 
$\tilde{\gamma}^{\rm gauge} = h$, 
the homogeneous part of the above equation can be cast 
into the following first order equation
for $(\tilde{\gamma}/h)'$:
\bdm
\left[\frac{h^{2} r^{2}}{w} 
\left(\frac{\tilde{\gamma}}{h} \right)' \right]' = 0 ,
\edm
with the solution $\tilde{\gamma} 
\propto h \int w/(r h)^{2}$.
The integration can be performed by using the relation
$[w/(r^2 h)]' = w/(r h)^{2}$, following from the background
equation (\ref{wh-eq-2}). Hence, the only non-gauge mode of
the homogeneous equations (\ref{Inh-Bog-odd}) is
\be
\tilde{\gamma}^{\rm hom} = \frac{w}{r^2},  \; \; \; 
\sqrt{2} \tilde{\alpha}^{\rm hom}_1 = h - \frac{1}{r} , \; \; \; 
\sqrt{2} \tilde{\alpha}^{\rm hom}_2 = \frac{w}{r} .
\label{hom-odd-sol}
\ee
(In order to verify that this solves the homogeneous part of 
Eqs. (\ref{Inh-Bog-odd}),
one uses again the first order equation (\ref{wh-eq-2})
for the background field $h$.)
We may finally use the two solutions 
$\tilde{\gamma}^{\rm hom}$
and $\tilde{\gamma}^{\rm gauge} = h$ 
to solve the inhomogeneous 
equation (\ref{gamma-1}) with source term
${\cal I} = -h \tilde{b} = -h (w-w^{-1})$:
\be
\tilde{\gamma}^{\rm inh} = 
\int \diff r \, \frac{{\cal I}}{r^2}
\left(\mu_{(1)} \tilde{\gamma}^{\rm gauge} +
\mu_{(2)} \tilde{\gamma}^{\rm hom} \right) ,
\label{inh-odd-gen}
\ee
with $\mu_{(1)} = \tilde{\gamma}^{\rm hom}/W$ and
$\mu_{(2)} = - \tilde{\gamma}^{\rm gauge}/W$, where
$W = \tilde{\gamma}^{\rm gauge}(\tilde{\gamma}^{\rm hom})' -
\tilde{\gamma}^{\rm hom}(\tilde{\gamma}^{\rm gauge})'$
is the Wronskian of the two homogeneous solutions.
A short computation yields $W = w/r^2$, 
and hence $\mu_{(1)} = 1$, $\mu_{(2)} = - r^{2} h / w$.
We thus end up with
\be
\tilde{\gamma}^{\rm inh} = \frac{w}{r^2}
\left[
\int \frac{r^{2} h}{w} \, \frac{h h'}{w} \, \diff r  - 
\frac{r^{2} h}{w} \int \frac{h h'}{w} \, \diff r  \right] .
\label{inhom-odd-sol}
\ee

This shows that the physical modes describing magnetic
perturbations with $J=1$ and odd parity form a two parameter
family. In particular,
the perturbations of the Higgs field become
\be
\delta H^{\rm odd} = \left( C_{1} \tilde{\gamma}^{\rm hom} +
C_{2} \tilde{\gamma}^{\rm inh} \right) .
\label{deltaH-res-odd}
\ee
(The arbitrary constant $C_2$ reflects the fact that the 
source terms $\delta B$ are themselves solutions to a 
homogeneous set of equations, implying that 
the inhomogeneity in Eq. (\ref{gamma-1}) 
is only fixed up to a multiplicative constant.)
Since the self-dual solution
$\tilde{\gamma}^{\rm hom}$ diverges like $1/r^{2}$ near the
origin, while the non self-dual part 
$\tilde{\gamma}^{\rm inh}$ diverges like $\int e^{r}/r$
at infinity, we conclude that there exist no small
magnetic perturbations of BPS monopoles and JZ dyons with
odd parity.

\subsection{Solutions to the $\delta \Phi$ equations}
\label{subsection-VI-C}

The electric perturbations  $\tilde{\phi}$ with odd parity
are governed by Eq. (\ref{el-odd}). Since the magnetic 
amplitude $\tilde{b}$ fulfills the same second order 
equation, we immediately conclude from the solution 
(\ref{sol-odd-B-eqs}) that
\be
\tilde{\phi}^{(1)} = w - \frac{1}{w}
\label{phi-1-odd}
\ee
solves Eq. (\ref{el-odd}). [In fact, using 
$(w \pm w^{-1})' = h (w \mp w^{-1})$, one has
$(w - w^{-1})''$ $=$ $h'(w + w^{-1}) + h^{2}(w - w^{-1})$
$=$ $[(w^{2}+1)/r^{2}+h^{2}](w - w^{-1})$.] 
The second solution is given by 
$\tilde{\phi}^{(2)} = \tilde{\phi}^{(1)} 
\int [\tilde{\phi}^{(1)}]^{-2} \diff r$. 
The integral can be carried out, and yields
\be
\tilde{\phi}^{(2)} = \frac{1}{r} \left(w + \frac{1}{w}
\right) - \frac{h}{w} .
\label{phi-2-odd}
\ee
[Using the background equations (\ref{wh-eq}) it is not 
hard to verify that this is indeed the second solution 
to Eq. (\ref{el-odd}).]
The electric perturbations with odd parity are, therefore,
\be
\delta \Phi^{\rm odd} = \frac{1}{r} 
\left( C_{1} \tilde{\phi}^{(1)} +
C_{2} \tilde{\phi}^{(2)} \right) ,
\label{deltaPhi-res-odd}
\ee
which remains finite for $r \rightarrow \infty$
only if $C_{1} = C_{2}$. However, as
$(\tilde{\phi}^{(1)} + \tilde{\phi}^{(2)})/r$ diverges 
like $1/r^{2}$ in the vicinity of the origin, we conclude 
that there exist no small electric perturbations of BPS 
monopoles and JZ dyons with odd parity.

\section{Even parity modes}
\label{section-VII}

\subsection{Solutions to the $\delta B$ equations}
\label{subsection-VII-A}

In order to solve Eqs. (\ref{even-B-eqs}), we first note that
the equation for $b'_{+}$ is a consequence of the remaining 
ones. Eliminating $b_{+}$ by using the last equation in
(\ref{even-B-eqs}), we obtain a system of three first order 
equations for $b_{-}$, $\beta_1$ and $\beta_2$. It is then
straightforward to decouple these equations, which yields a
third order equation for $b_{-}$. Since $b_{-}$ enters this 
equation only via its derivatives, one concludes that 
$b_{-} = const.$ is a solution. In fact, one easily verifies
that (any constant times)
\be
b^{(0)}_{-} = 2, \; \; b^{(0)}_{+} = \sqrt{2} 
\left(w + \frac{1}{w}\right), \; \; 
\beta^{(0)}_{1} = -h, \; \; \beta^{(0)}_{2} = \frac{h}{w}
\label{Sigma-0}
\ee
solves Eqs. (\ref{even-B-eqs}). In order to find the 
remaining two solutions, it is convenient to define the 
quantities
\bea
\Sigma & = & h \, b_{-} + 
\left( \beta_{1} - w \beta_{2} \right) ,
\nonumber\\
\Delta & = & h^{-1} \left( \beta_{1} + w \beta_{2} \right) .
\label{Sigma-def}
\eea
Since $\Sigma$ and $\Delta$ vanish for the solution 
(\ref{Sigma-0}), it is possible to derive a system of two 
first order equations for these quantities. Using the three 
equations for $b_{-}$, $\beta_{1}$ and $\beta_{2}$, one 
finds after some manipulations $\Sigma' = 2 h^{2} \Delta$ 
and $\Delta' = h^{2} r^{2}(w^{2}+1)^{-1}\Sigma$, which also 
yields the following second order equation for $\Sigma$:
\be
\Sigma'' - 2 \/\frac{h'}{h} \/ \Sigma' -
2 \/ \frac{w^{2}+1}{r^{2}} \/ \Sigma = 0.
\label{Sig2nd}
\ee
Considering the quantity $\Sigma / h$, we obtain 
a Schr\"odinger equation with potential $P(r)$,
\be
(\Sigma / h)'' = P(r) (\Sigma / h) , \; \; \; 
\mbox{where} \; P(r) = 2 \, \frac{w^{2}+1}{r^{2}} + 
\frac{(h^{-1})''}{h^{-1}} .
\label{Sigma-eq}
\ee
Having solved this equation, one obtains the magnetic
amplitude $b_{-}$ from the definition of $\Sigma$ and the
first equation in (\ref{even-B-eqs}). In order to find
$b_{+}$ one uses the last equation in (\ref{even-B-eqs})
and solves Eqs. (\ref{Sigma-def}) for 
$\beta_{1}$ and $\beta_{2}$. This yields
\bea
b_{-} & = & \int \frac{\Sigma'}{h} \, \diff r,
\nonumber\\
\sqrt{2} b_{+} & = & \frac{1}{h} 
\left(w + \frac{1}{w} \right) \left(hb_{-} - \Sigma \right) - 
\frac{1}{2h^{2}} \left(w - \frac{1}{w} \right) \Sigma' .
\label{bpm}
\eea

The formulas for $\beta_{1}$ and $\beta_{2}$ in terms of 
$\Sigma$ are not given here, since $\delta B$ 
can be expressed in terms of $b_{-}$ and $b_{+}$ alone. 
This is seen as follows:
Using the relations (\ref{A13}), the terms tangential to 
$S^{2}$ in the expansion (\ref{exp-B}) for 
$\delta B^{\rm even}$ can be written in the form
\bea
\beta_1 \, \taur \diff K + \beta_2 \, K \diff \taur & = &  
\frac{1}{w^{2}-1} \left[
\left(w \beta_{2} - \beta_{1} \right) \hatDiff X + \sqrt{2}
\left(\beta_{2} - w \beta_{1} \right) \hatDiff Y \right]
\nonumber\\ & = &
\frac{1}{w^{2}-1} \left[
\left(h \, b_{-} - \Sigma \right) \hatDiff X +
\left(h \, b_{+} \right) \hatDiff Y \right] ,
\nonumber
\eea
where we have used the last equation in (\ref{even-B-eqs})
and the definition (\ref{Sigma-def}) to get rid of
$\beta_{1}$ and $\beta_{2}$. Now using Eq. (\ref{bpm})
for $b_{-}$, we have $\Sigma' = h b'_{-}$, and thus
$(h b_{-} - \Sigma)' = h' b_{-}$,
which enables us to write the term
proportional to $X$ in Eq. (\ref{exp-B}) as
$r^{-2} b_{-} \diff r$ $=$ $(w^{2}-1)^{-1}h'b_{-} \diff r$
$=$ $(w^{2}-1)^{-1} \diff (h b_{-} - \Sigma)$. 
Hence, the terms proportional to $X$ and $\hatDiff X$
combine to an exact covariant derivative, which
finally yields the result 
\be
\delta B^{\rm even} = \frac{1}{w^{2}-1}
\left\{ \Diff \left[ \left(h b_{-} - \Sigma \right) X
\right] \, + \, b_{+} \Diff \left[ h Y \right] \right\} .
\label{delB-res}
\ee

This shows that all three magnetic modes with even parity 
are obtained from Eqs. (\ref{Sigma-eq}), (\ref{bpm}) 
and (\ref{delB-res}). In particular, the trivial solution 
$\Sigma^{(0)} = 0$ of Eq. (\ref{Sigma-eq}) gives rise to
the solution (\ref{Sigma-0}), which yields
\be
\delta B^{{\rm even}, (0)} = \frac{2}{w^{2}-1}
\left\{ \Diff \left( h X \right) + \frac{1}{\sqrt{2}}
\left(w + \frac{1}{w} \right) 
\Diff \left( h Y \right) \right\} .
\label{delB-res-0}
\ee

The two non-trivial solutions of Eq. (\ref{Sigma-eq}) are
not (yet) known in closed form. However, their qualitative 
behavior can be discussed rigorously: The potential $P(r)$ 
is positive definite for all finite values of $r$. 
(In fact, using the background equations to compute 
$(h^{-1})''$, one finds from Eq. (\ref{Sigma-eq})
\be
P(r) = \frac{1}{r^{2}} + \left(\frac{h'}{h}\right)^{2} +
\left( \frac{1}{r} + \frac{h'}{h} \right)^{2} ,
\label{potential}
\ee
which is manifestly non-negative, and vanishes only for
$r \rightarrow \infty$.) 
As $h'/h = 1/r - 2r/15 + {\cal O}(r^{3})$
in the vicinity of the origin, we have 
$P(r) = 6 / r^{2} + {\cal O}(1)$, implying that the 
fundamental solutions behave like 
$\Sigma^{(1)}/h \propto 1/r^{2}$ and
$\Sigma^{(2)}/h \propto r^{3}$. 
For $r \rightarrow \infty$ one has 
$P(r) = 2 / r^{2} + {\cal O}(1/r^{3})$, which yields
$\Sigma^{(1)}/h \propto 1/r$ and
$\Sigma^{(2)}/h \propto r^{2}$. The 
monotonicity property of $\Sigma/h$, following from the
positivity of the potential (\ref{potential}), 
enables one to conclude that
the solution which diverges at the origin
remains finite at infinity, and vice-versa. (Also we have
used global existence, following from the linearity 
of Eq. (\ref{Sigma-eq}) and from the finiteness of the 
potential for $r \neq 0$.). Since $h$ behaves like $r$ near 
the origin and approaches the constant value $-1$ at
infinity, the two non-trivial solutions of 
Eq. (\ref{Sigma-eq}) behave as follows:
\be
\Sigma^{(1)} \propto r^{-1} , \; \; \; 
\Sigma^{(2)} \propto r^{4} , \; \; 
\mbox{for $r \rightarrow 0$} ,
\label{nearzero}
\ee
and
\be
\Sigma^{(1)} \propto r^{-1} , \; \; \; 
\Sigma^{(2)} \propto r^{2} , \; \; 
\mbox{for $r \rightarrow \infty$} .
\label{nearinfty}
\ee

In Sect. \ref{subsection-IV-C} we have shown that the 
angular momentum can be expressed as a boundary
integral. The relevant quantity appearing in Eq. (\ref{JMG})
is the difference of $(b_{-} + \sqrt{2} w b_{+})$
between infinity and zero. For the solution
$\Sigma^{(1)}$ this quantity remains finite at infinity,
whereas it diverges logarithmically at the origin.
(Note that the leading power, $1/r^{2}$, cancels in the
above combination.) On the other hand, 
$(b_{-} + \sqrt{2} w b_{+})$ remains bounded at the origin 
for the solution $\Sigma^{(2)}$, 
whereas it obviously diverges like
$r^{2}$ at infinity. Hence, neither of the two non-trivial
solutions to Eq. (\ref{Sigma-eq}) gives rise to a finite
angular momentum. The fact that we are able to decide this 
without solving the inhomogeneous equations 
(\ref{Inh-Bog-even}) for $\delta H$ and $\delta A$ 
follows from the observation that the 
angular momentum depends on the magnetic perturbations only 
via the field $\delta B$. Moreover, only the 
boundary values of the gauge invariant quantity 
$\delta B$ are needed to obtain the 
magnetic contribution to the total angular momentum.
 
Surprisingly enough, the third solution,
given in closed form in Eq. (\ref{Sigma-0}),
does give rise to a finite angular momentum, although the 
amplitude $b_{+}$ diverges at infinity.
By virtue of Eq. (\ref{JMG}) we have
\be
\delta J^{\rm mg} = -\frac{\pi}{3} 
\left[ b_{-} + \sqrt{2} w b_{+} \right]_{0}^{\infty} =
\frac{2 \pi}{3} \, . 
\label{JMG-finite}
\ee
In order to decide whether this is an acceptable 
perturbation, we have to compute the physical fields 
$\delta H$ and $\delta A$.

\subsection{Solutions to the inhomogeneous BPS equations}
\label{subsection-VII-B}

We now discuss the inhomogeneous linearized 
Bogomol'nyi equations (\ref{Inh-Bog-even}). These can be
written as a fourth order equation for either of the
four variables $\gamma_{\pm}$, $\alpha_{1,2}$, 
parametrizing $\delta H^{\rm even}$ and 
$\delta A^{\rm even}$.
Since $\gamma_{-}$ vanishes for the residual gauge 
mode of the system (\ref{Inh-Bog-even}), 
the equation for $\gamma_{-}$ is
only of third, rather than fourth order.
One finds
\be
\left(r^{2} \gamma'_{-}\right)'' - 2 h
\left(r^{2} \gamma'_{-}\right)' - 2 \left(1+2w^{2}\right)
\gamma'_{-}+ 4 h \left(1-w^{2}\right) \gamma_{-} = 
{\cal I}[\delta B^{\rm even}] ,
\label{gamma-eq}
\ee
where the inhomogeneity is an expression
in terms of $b_{\pm}$ and $\beta_{1,2}$. Using
Eqs. (\ref{even-B-eqs}) for these amplitudes, 
it is possible to express ${\cal I}[\delta B^{\rm even}]$
in terms of $b_{+}$ and $b_{-}$ alone:
\be
{\cal I}[\delta B^{\rm even}] = \frac{2w}{1-w^{2}} \left[w 
\left(h b_{-}\right)' + \sqrt{2}
\left(h b_{+}\right)' \right] .
\label{IdeltaB}
\ee

The solutions to the homogeneous problem, that is, the
solutions to the linearized Bogomol'nyi equations are
known in closed form \cite{Mottola78},
\cite{Adler79}, \cite{Akhoury80}.
In fact, they can be expressed
in terms of the quantities $w$ and $h$.
Using the background relations
(\ref{wh-eq}), it is not difficult -- and not 
particularly pleasant either -- to verify that
the three solutions of the homogeneous 
Eq. (\ref{gamma-eq}) are
\bea
\gamma^{(1)}_{-} & = & h' = \frac{1}{r^{2}}
\left(w^{2}-1 \right) ,
\nonumber\\
\gamma^{(2)}_{-} & = & \frac{1}{r^{3}} \left[
w^{2} - \left(rh-1\right)\right] ,
\nonumber\\
\gamma^{(3)}_{-} & = & 
w^{2} + 2 \, \left(rh-1\right) .
\label{hom-1-3}
\eea
Concerning these solutions of the homogeneous linearized 
PBS equations, we note the following.
\begin{itemize}
\item
The amplitude $\gamma^{(2)}_{-}$ is not finite at 
the origin, while $\gamma^{(3)}_{-}$ becomes unbounded 
at infinity. Hence,
neither $\gamma^{(2)}_{-}$ nor $\gamma^{(3)}_{-}$ give
rise to small perturbations of $\delta H^{\rm even}$. 
\item
The third solution, $\gamma^{(1)}_{-}$, does give rise 
to an acceptable physical mode. The latter corresponds 
to a translation along the $z$-axis. 
This is seen by differentiating
the background field $H = \taur h$ with respect to
$\partial_{z} = \cos\!\vartheta \partial_{r} - 
r^{-1} \sin\!\vartheta \partial_{\vartheta}$, which 
yields $\partial_{z} H = h' X + r^{-1} h \sqrt{2} Y$,
and hence $\gamma_{-} = h'$.
(Note that the coefficient in front of $Y$ 
is not the amplitude $\gamma_{+}$ introduced in Eq. 
(\ref{delta-H}), since the latter was defined in a 
gauge where $\delta A$ is tangential to $S^{2}$; 
see Eq. (\ref{delta-A}). The only quantity which is not 
affected by the corresponding gauge transformation 
is $\gamma_{-}$; see also Appendix \ref{AA-4}.)
\item
The remaining two physical zero modes in the sector 
$J=1$ correspond to translations along the 
$x$- and $y$-axis. 
They do not occur in the above calculation, since, 
for reasons of symmetry, we have restricted the
harmonic decompositions to the magnetic quantum 
number $M=0$.
\item
We recall that none of the above solutions to
the linearized BPS equations can contribute to 
the angular momentum, since only the
field $\delta B$, describing the non self-dual
perturbations, enters the expression (\ref{T-mg-2}).
\end{itemize}

It remains to find the solutions to the inhomogeneous
equation (\ref{gamma-eq}). Since all solutions to 
the homogeneous problem are known, we can apply 
standard methods to obtain the particular solution
$\gamma_{-}^{\rm inh}$. 
For given inhomogeneity ${\cal I}$ one has
\be
\gamma_{-}^{\rm inh} = \sum \gamma_{-}^{(k)}
\int \mu_{(k)} \, 
\frac{{\cal I}}{r^2} \, \diff r ,
\label{inh-gen}
\ee
where the three quantities $\mu_{(k)}$ are obtained 
from the homogeneous solutions $\gamma^{(k)}_{-}$ by
\be
\mu_{(k)} = \frac{
\varepsilon_{ijk} \,
\gamma^{(i)}_{-} (\gamma^{(j)}_{-})'}{
\varepsilon_{m n \ell} \,
\gamma^{(m)}_{-} (\gamma^{(n)}_{-})' 
(\gamma^{(\ell)}_{-})'' } \, .
\label{inh-gen-2}
\ee
A rather lengthy computation yields the value
$-24 w^{2} / r^{4}$ for the Wronskian in the 
denominator, and then 
\be
\mu_{(1)} = \frac{1}{4}
\left(1 + r^{2} + \sinh^{2}(r) \right) , 
\; \; \; 
\mu_{(2)} = \frac{1}{4}
\left( \sinh(r)\cosh(r) - r - 
\frac{2}{3} r^{3} \right) , \; \; \; 
\mu_{(3)} = -\frac{1}{12} .
\label{mu-1-3}
\ee
In the previous section we have argued that only 
the solution (\ref{Sigma-0}) of the equations 
(\ref{even-B-eqs}) for $\delta B^{\rm even}$ 
gives rise to a finite angular momentum. 
Hence, it remains to compute $\gamma_{-}^{\rm inh}$ 
for the source term given by $b^{(0)}_{-} = 2$ and 
$b^{(0)}_{+} = \sqrt{2}(w + w^{-1})$.
Using the expression (\ref{IdeltaB}), 
we immediately have
\be
{\cal I}[\delta B^{{\rm even}(0)}] = 
-4 \left(
h^{2} + \frac{2 w^{2}+1}{r^{2}} \right) .
\label{IdeltaB-0}
\ee
The solution $\gamma^{\rm inh}_{-}$ is now obtained 
from Eqs. (\ref{inh-gen}), (\ref{inh-gen-2}), 
(\ref{mu-1-3}) and (\ref{IdeltaB-0}). An expansion in
powers of $r$ reveals that $\gamma^{\rm inh}_{-}$ 
diverges like $1/r$ in the vicinity of the origin. 
Since $\gamma^{(2)}_{-}$ diverges like $1/r^3$, while  
$\gamma^{(1)}_{-}$ and $\gamma^{(3)}_{-}$
are well-behaved for $r \rightarrow 0$,
there is no combination of $\gamma_{-}^{\rm inh}$
with a homogeneous solution (\ref{hom-1-3})
which remains bounded at the origin. 
Hence, we conclude that there exist no bounded 
magnetic perturbations $\delta H$, $\delta A$, 
which give rise to finite angular momentum.

\subsection{Solutions to the $\delta \Phi$ equations}
\label{subsection-VII-C}

It remains to discuss the electric perturbations with 
even parity. The latter are governed by 
Eqs. (\ref{el-even}) for $\phi_{-}$ and $\phi_{+}$. 
We have already argued that three solutions of these 
equations coincide with the magnetic solutions 
$b^{(0)}_{\pm}$, $b^{(1)}_{\pm}$ and $b^{(2)}_{\pm}$, 
discussed in Sect. \ref{subsection-VII-A}.
In order to find the remaining solution, 
it is convenient to write the system (\ref{el-even}) 
in the form of a second order equation with two 
inhomogeneities. The manipulations by which this can 
be achieved are discussed in Appendix \ref{AA-5}. 
Introducing the quantity 
$\tilde{\Sigma}$ in the same way as in the
magnetic case [see Eq. (\ref{bpm})],
\be
\tilde{\Sigma}' \, = \, h \phi'_{-} ,
\label{def-tildesigma}
\ee
one eventually finds the equation
\be
\tilde{\Sigma}'' - 2 \/\frac{h'}{h} \/ \tilde{\Sigma}' - 
2 \/ \frac{w^{2}+1}{r^{2}} \/ \tilde{\Sigma} = 
-k_{3} \, h' - k_{0} \, 2 \, \frac{w^{2}+1}{r^{2}} ,
\label{Sig2nd-inh}
\ee
where $k_{0}$ and $k_{3}$ are integration constants. 
The homogeneous part of this equation coincides with 
the corresponding magnetic equation (\ref{Sig2nd}). 

The particular solution for $k_{3} = 0$ is 
$\tilde{\Sigma} = k_{0} \tilde{\Sigma}^{(0)}$, with
$\tilde{\Sigma}^{(0)} = 1$. This yields
$\phi^{(0)}_{-} = const.$, which coincides with the
magnetic solution (\ref{Sigma-0}).
The two solutions to the homogeneous problem,
$\tilde{\Sigma}^{(1)}$ and $\tilde{\Sigma}^{(2)}$, say,
coincide with the two remaining magnetic solutions 
obtained from the homogeneous equation (\ref{Sig2nd}).
The additional solution, 
$\tilde{\Sigma} = k_{3} \tilde{\Sigma}^{(3)}$, 
which is not present in the 
magnetic case, is the particular solution for
$k_{0} = 0$. Again, this can be given in closed form by 
introducing the quantity $S = \tilde{\Sigma}/h^{2}$.
Using again the background equation (\ref{wh-eq-2}) 
for $h$, a short computation shows that the LHS of 
Eq. (\ref{Sig2nd-inh}) assumes the form 
$(S'h^{2})'+2h'S$. Hence, the desired solution is
$S = -k_{3}/2$, that is, 
$\tilde{\Sigma}^{(3)}=-h^{2}/2$. By virtue of 
Eq. (\ref{def-tildesigma}) this yields 
$\phi^{(3)}_{-} \propto h$.

We thus conclude that the four electric perturbations 
with even parity are given by
\bea
\phi^{(0)}_{-} & = & 2 , \; \; \; 
\phi^{(0)}_{+} =  \sqrt{2} 
\left( w + \frac{1}{w} \right) , 
\label{phi-0}\\
\phi^{(3)}_{-} & = & 2 h , \; \; \; 
\phi^{(3)}_{+}  =  \frac{\sqrt{2}}{w} 
\left( r \, h \right)' ,
\label{phi-3}\\
\phi^{(1,2)}_{-} & = & \int \diff r \, 
\frac{(\tilde{\Sigma}^{(1,2)})'}{h} , \; \; \; 
\phi^{(1,2)}_{+} = \frac{1}{\sqrt{2}} \left[
\left(w + \frac{1}{w} \right) \phi^{(1,2)}_{-} - 
\frac{r^{2} (\phi_{-}^{(1,2)})''}{2 \, w} \right] ,
\label{phi-12} 
\eea
where $\tilde{\Sigma}^{(1,2)}$ are the two non-trivial 
solutions to $\tilde{\Sigma}'' - 2 (h'/h) \tilde{\Sigma}' - 
2 r^{-2} (w^{2}+1) \tilde{\Sigma} = 0$.
The angular momentum is obtained 
from Eq. (\ref{JEL}) and the above solutions. One finds
\bea
\delta J^{{\rm el}(0)} & = &
\frac{2 \pi}{3} \left[r^{2} h' 
\, (1-rh)\right]^{\infty}_{0} ,
\nonumber\\
\delta J^{{\rm el}(3)} & = &
\frac{2 \pi}{3} \left[ rh 
\, (h-rh') \right]^{\infty}_{0} ,
\nonumber\\
\delta J^{{\rm el}(1,2)} & = &
-\frac{\pi}{3} \left[ r^{2}h 
\,  \left( \frac{h'}{h} (rh-1) 
\phi^{(1,2)}_{-} + (\phi^{(1,2)}_{-})' - 
\frac{r}{2} (\phi^{(1,2)}_{-})'' \right) 
\right]^{\infty}_{0} .
\label{JEL-0-3}
\eea

It is easy to see that the only combination of 
$\phi^{(0)}$ and $\phi^{(3)}$ which gives rise 
to finite angular momentum is their sum.
However, since the amplitudes entering 
$\delta \Phi$ are
$\phi_{+}/r$ and $\phi_{-}/r$ [see Eq. (\ref{exp-Phi})], 
the perturbation $\delta \Phi$ obtained from 
$\phi^{(0)}+\phi^{(3)}$ diverges 
at the origin like $1/r$.

The behavior of $\tilde{\Sigma}^{(1)}$ given in 
Eqs. (\ref{nearzero}) and (\ref{nearinfty})
implies that $\phi^{(1)}_{-} = {\cal O}(1/r^{2})$
as $r\rightarrow 0$, and 
$\phi^{(1)}_{-} = {\cal O}(1/r)$ as 
$r\rightarrow \infty$. Using this in 
the above expression shows that the angular 
momentum is again finite,
$\mid\delta J^{{\rm el}(1)}\mid < \infty$.
As above, the perturbation $\delta \Phi$ 
is, however, not bounded at the origin.

The solution $\phi^{(2)}_{-}$ diverges like $r^{2}$ 
as $r\rightarrow \infty$. However, the leading order
terms in the expression for $\delta J^{{\rm el}(2)}$ 
cancel, and so do the next-to-leading order terms. 
Hence, like $\delta J^{{\rm el}(0)}$ and 
$\delta J^{{\rm el}(3)}$, $\delta J^{{\rm el}(2)}$ 
diverges only with the first power of $r$, implying 
that there exist linear combinations of 
$\phi^{(2)}_{-}$ with $\phi^{(0)}_{-}$ (or 
$\phi^{(3)}_{-}$) which give rise to finite angular 
momentum. Since $\phi^{(0)}_{-}/r$ is not bounded at 
the origin, while $\phi^{(2)}_{-}/r$ is bounded, 
only linear combinations of $\phi^{(2)}_{-}$
with $\phi^{(3)}_{-}$ need to be considered.
However, the latter give rise to perturbations 
$\delta \Phi$ which are not bounded at infinity. 
(Note that $\phi^{(2)}_{+}/r$ behaves like $e^r/r$, 
whereas $\phi^{(3)}_{+}/r$ grows like $e^r$.) 

We therefor conclude that all electric perturbations 
of BPS monopoles and JZ dyons which give rise 
to finite angular momentum are either
unbounded at the origin or at infinity.

\section{Conclusions}

In this paper we have presented a gauge invariant
approach to the stationary perturbations of Julia-Zee 
dyons and BPS monopoles. Restricting our attention to 
axisymmetric perturbations, we have found
three sets of modes in each parity sector.
\begin{itemize}
\item
Electric perturbations: These are manifestly gauge 
invariant, since the electric background field 
vanishes after a hyperbolic rotation. There exist 
even parity perturbations with finite angular momentum; 
however, these are either not well-behaved at the 
origin or at infinity. The same is true of the 
odd parity perturbations, which give
rise to axial deformations.
\item
Non self-dual magnetic perturbations:
These are perturbations which satisfy the linearized
field equations, but are not at the same time subject to
the linearized Bogomol'nyi equations. As the corresponding
background field vanishes, the non self-dual magnetic perturbations
are also gauge invariant. Like in the 
electric case, there exist even parity perturbations
with finite angular momentum. However,
neither these nor the odd parity modes are well behaved.
\item
Self-dual magnetic perturbations: These have been 
investigated before and are known to be physically 
not acceptable, apart from the translational modes. 
Moreover, general considerations show that self-dual
modes cannot contribute to the angular momentum.
The fact that all solutions to the linearized BPS 
equations are known in closed form is, however, 
very useful to reconstruct the physical fields
$\delta H$ and $\delta A$ for the non self-dual modes.
\end{itemize}
In conclusion, we would like to emphasize that the
distinguished properties of the BPS background are very 
critical to the methods developed in this paper. 
Whether the main result -- the fact that there exist no 
rotational excitations of Julia-Zee dyons and BPS 
monopoles -- generalizes to more general background 
configurations is an open problem. In particular, the 
effect of a Higgs potential and the coupling 
to gravity need to be investigated for the excitations 
of Julia-Zee dyons.

\appendix
\section{BPS background}
\label{AA-1}

The appropriate way to treat axial perturbations 
in gauge theories is by using the isospin harmonics
introduced in Sect. \ref{subsection-IV-A}.
It is, therefore, suited to write the background fields
in terms of the spherical su(2) basis 
$\taur$, $\tau_{\vartheta}$, $\tau_{\varphi}$, 
defined by
\be
\taur = \taubold \cdot \hat{\radbold} , \; \; \; 
\diff\taur = \tautheta \diff\vartheta + 
\tauphi \sin \! \vartheta \diff\varphi , 
\label{A1}
\ee
where $\hat{\radbold} \equiv \radbold / r$ is the radial 
unit direction, and $\taubold = \sigbold/(2i)$.
The commutation relations of the Pauli matrices imply
$[\taur , \tautheta] = \tauphi$ (and cyclic), from which
one obtains the formulas
\be
[\taur , \diff \taur] = -\hatast \diff \taur, \; \; \; 
[\diff \taur , \hatast \diff \taur] = 0 ,
\nonumber
\ee
\be
[\diff \taur , \diff \taur] =
[\hatast \diff \taur , \hatast \diff \taur] = 
2 \, \taur \, \diff \Omega ,
\label{A2}
\ee
where $\hatast$ denotes the Hodge dual with respect to 
the standard metric of the two-sphere $S^{2}$.
It is helpful to recall that the radial unit direction 
$\hat{\radbold}$ is a vector valued eigenfunction of 
the spherical Laplacian with eigenvalue $2$, implying 
\be
\diff \hatast \diff \taur = - 2 \, 
\taur \, \diff \Omega \, .
\label{A3}
\ee

In terms of $\taur$, the ``Witten ansatz'' for the 
spherically symmetric connection one-form
assumes the simple form
\be
A = [1-w(r)] \, \hatast \diff \taur  ,
\label{A4}
\ee
since
$\hatast \diff \taur = r^{-2} (\radbold \times \taubold)  
\cdot \diff \radbold$. 
Using the commutation relations (\ref{A2}),
the gauge covariant derivatives of 
$\taur$, $\diff \taur$ and $\hatast \diff \taur$ become
\be
\Diff \taur  =  w \, \diff \taur , \; \; \;
\Diff \diff \taur = 0 ,
\nonumber
\ee
\be
\Diff \hatast \diff \taur  =  -2 w \, \taur \diff \Omega .
\label{A5}
\ee
The Bogomol'nyi equations, $\ast F = \Diff H$, 
can easily be written out by using Eqs. (\ref{A5}), 
the ansatz $H = h(r) \taur$, and the formulas 
$\ast (\diff r \we \hatast \diff \taur) = -\diff \taur$, 
$\ast \diff \Omega = \diff r/r^{2}$ for the
three-dimensional Hodge dual. One finds
\be
\ast F  =  w' \, \diff \taur + 
\frac{w^{2}-1}{r^{2}} \, \taur \diff r ,
\nonumber
\ee
\be
\Diff H  =   h w \, \diff \taur + h' \, \taur \diff r ,
\label{A6}
\ee
which yields the well-known first order equations 
(\ref{wh-eq}) for $w(r)$ and $h(r)$.

\section{The $2+1$ decomposition}
\label{AA-2}

The axial perturbation equations for a static, spherically
symmetric su(2) valued function involve the three-dimensional
gauge covariant Laplacian with respect to the gauge potential
(\ref{A4}). As the latter is tangential to $S^2$, 
the three-dimensional gauge covariant derivative operator is
\be
\Diff = \diff r \we \partial_{r} + \hatDiff ,
\label{A7}
\ee
where
\be
\hatDiff = \hat{\diff} \, \cdot \, + [A, \cdot \, ] , \; \; 
\mbox{with $\hat{\diff} = 
\diff \vartheta \we \partial_{\vartheta}+
\diff \varphi \we \partial_{\varphi}$} \, .
\label{A12}
\ee
For an arbitrary Lie algebra valued function $f$ we thus have
\be
\ast \Diff \ast \Diff \, 
\left( \frac{f}{r} \right) = \frac{1}{r} 
\left( \partial_r^2 + \frac{1}{r^2}
\hatast \hatDiff \hatast \hatDiff \right) f ,
\label{A8}
\ee
where the factor $1/r$ is introduced for convenience.
(Here we have used
$\ast \diff r = r^{2} \diff \Omega$ and 
$\ast \hatDiff f = -\diff r \we \hatast \hatDiff f$.)
The above formula enables us to immediately write down
the $2+1$ decomposition of the electric perturbation 
equation (\ref{deltaphi}).
With $f = r \delta \Phi$ this becomes
\be
\left( \partial_r^2 + \frac{1}{r^2}
\hatast \hatDiff \hatast \hatDiff \right) 
\left( r \delta \Phi \right) = -
[H,[H, r \delta \Phi] \, ] \, .
\label{A9}
\ee

The $2+1$ decomposition of the (first order) 
magnetic equations
(\ref{deltab}) with respect to the ansatz
\be
\delta B = \frac{1}{r^2} \, b \, \diff r + \hat{B} 
\label{A10}
\ee
was given in Sect. \ref{subsection-V-B}; 
see Eqs. (\ref{BbB-eqs}).
(Here $\hat{B}$ denotes an su(2) valued one-form
tangential to $S^2$, and $b$ is an su(2)
valued scalar field.) We owe the proof of
the assertion that $b$ is subject to the same 
second order equation as the 
scalar electric perturbation $\delta \Phi$.
In order to see this, one applies
$\hatast \hatDiff \hatast$ on the second, 
and $\partial_r$ on the third 
equation in (\ref{BbB-eqs}). A short 
calculation yields
\bdm
\left(\partial_r^2 + \frac{1}{r^2} 
\hatast \hatDiff \hatast \hatDiff 
\right) b = -\hatast 
[(\hatDiff H - \hatast A') \, ,  
\hat{B}] - [H,\hatast \hatDiff \hat{B}] .
\edm
The $2\!+\!1$ decomposition of the Bogomol'nyi 
equation gives $\hatDiff H = \hatast A'$, 
implying that the first commutator on the RHS 
vanishes. By virtue of the first equation in 
(\ref{BbB-eqs}), the second commutator becomes 
$[H,[H,b]]$, which yields the result
\be
\left( \partial_r^2 + \frac{1}{r^2}
\hatast \hatDiff \hatast \hatDiff \right) \, b = -
[H,[H, b] \, ] \, .
\label{A11}
\ee
Hence, the equation (\ref{A9}) for the scalar
electric perturbation, $r \delta \Phi$, 
coincides with the second order equation 
(\ref{A11}) for the scalar part of the magnetic 
perturbation, $b \equiv r^2 (dr,B)$.

\section{Harmonic analysis}
\label{AA-3}

By virtue of the above decompositions, the task of 
writing out the perturbation equations 
reduces to the problem of 
computing the gauge covariant derivative $\hatDiff$
of su(2) valued functions and one-forms over $S^2$.
We have already argued in Sect. \ref{subsection-IV-A} 
that the $J=1$ sector is spanned by the three scalar 
harmonics $X$, $Y$ and $Z$, defined in terms of 
$\taur$ and $K \equiv \cos\!\vartheta$ 
[see Eq. (\ref{XYZ})], and the four one-forms 
$\diff X= -\sqrt{2}\diff Y$, $\hatast \diff X$, 
$\diff Z$ and $\hatast \diff Z$ [see Eq. (\ref{dXdYdZ})].
Instead of the latter, it is very convenient to use the 
linear combinations $\taur \diff K$, $K \diff \taur$
and their duals. The entire harmonic decomposition
is then obtained form the formulas
\bea
\hatDiff X & = & \taur \, 
\diff K + w \, K \, \diff \taur ,
\nonumber\\
\sqrt{2} \, \hatDiff Y & = & -w \, 
\taur \, \diff K - K \, \diff \taur ,
\nonumber\\
\sqrt{2} \, \hatDiff Z & = & w \, 
\taur \hatast \diff K - K \hatast \diff \taur
\label{A13} 
\eea
for the covariant derivatives of the scalar basis, and
the relations
\bea
\hatast \hatDiff (\taur \diff K) & = &
-w \, \sqrt{2} Z ,
\nonumber\\
\hatast \hatDiff (K  \diff\taur) & = &
\sqrt{2} Z ,
\nonumber\\
\hatast \hatDiff (\taur \hatast \diff K) & = &
w \, \sqrt{2} Y - 2 X ,
\nonumber\\
\hatast \hatDiff (K \hatast \diff\taur) & = &
\sqrt{2} Y - 2 w \, X 
\label{A14}
\eea
for the covariant derivatives of the basis 
one-forms. (The equation for $\hatDiff X$ and 
Eqs. (\ref{A14}) are immediate consequences of
Eq. (\ref{A5}), while the derivations of the 
expressions for $\hatDiff Y$ and $\hatDiff Z$ 
require slightly more work.)
 
As an illustration we compute
$\hatast \hatDiff \hatast \hatDiff 
\left( r \delta \Phi \right)$, where we
use the expansion (\ref{exp-Phi}) to write
$r\delta \Phi = \phi_{-}X + \phi_{+}Y + \tilde{\phi}Z$.
For the first term we find, for instance
\bea
\hatast \hatDiff \hatast \hatDiff 
\left( \phi_{-} X \right) & = & \phi_{-} \hatast \hatDiff
\left[ \taur \hatast \diff K + w \, K \hatast 
\diff \taur \right] 
\nonumber\\
& = & \left[ 2 \sqrt{2} w \, Y - 2 \left(w^{2}+1 \right)
\right] \, \phi_{-} .
\nonumber
\eea
A similar computation for the second and third term 
gives the result
\bea
\hatast \hatDiff \hatast \hatDiff 
\left( r\delta \Phi \right) 
& = & \left[-2 \, (w^{2}+1)  \phi_{-} + 
2\sqrt{2} \, w \phi_{+} \right] X
\nonumber\\ & + &
\left[2\sqrt{2} \, w \phi_{-} - 
(w^{2}+1) \phi_{+} \right] Y
\nonumber\\ & - &
\left[(w^{2}+1) \tilde{\phi} \right] Z,
\label{A15}
\eea
which, together with the $2+1$ decomposition 
formula (\ref{A9}) and 
$[H,[H,(r\delta \Phi)] = -(\phi_{+} Y + \tilde{\phi} Z)$, 
yields the desired perturbation equations (\ref{el-even}) 
and (\ref{el-odd}).

\section{Gauge transformations}
\label{AA-4}

In this Appendix we show that there 
exists a gauge for which
the perturbations $\delta H$ and 
$\delta A$ assume the expansions 
(\ref{delta-H}) and (\ref{delta-A}), respectively. 
We also establish that the coefficients
are gauge invariant, up to the residual gauge 
transformations given in Eqs. (\ref{res-even}) 
and (\ref{res-odd}). For simplicity, we focus 
on the even parity sector; the manipulations
for the odd parity sector are completely analogous.
The general expansions for $\delta H^{\rm even}$ and 
$\delta A^{\rm even}$ are
\be
\delta H^{\rm even}  =  
\bar{\gamma}_{-} X + \bar{\gamma}_{+} Y ,
\label{A16}
\ee
\be
\delta A^{\rm even}  =  
\bar{\alpha}_{0} \, Z \, \diff r + 
\bar{\alpha}_1 \, \taur \hatast \diff K + 
\bar{\alpha}_2 \,
K \hatast \diff \taur ,
\label{A17}
\ee
where the bars have been introduced to tell the 
amplitudes apart from the ones introduced in 
Eqs. (\ref{delta-H}) and (\ref{delta-A}).
Under a gauge transformation with an su(2) valued
function $\chi$ one has
\bea
\delta H & \rightarrow & \delta H + [H, \chi] ,
\nonumber\\
\delta A & \rightarrow & \delta A + \Diff \chi ,
\label{A18}
\eea
where, as usual, $H$ is the background Higgs field 
and $\Diff$ the covariant derivative with 
respect to the background potential $A$. 
The strategy is to write $\delta H$ and $\delta A$ 
as sums of a pure gauge and an (almost) gauge 
invariant part. For $\delta A$ this is achieved 
by a partial integration of the radial part, 
and by using the expressions (\ref{A13}) for
the covariant derivatives of the isospin basis.
The radial part of $\delta A^{\rm even}$ 
can be written as
\bdm
\bar{\alpha}_{0} \, Z \, \diff r = \Diff 
\left[ Z \int \bar{\alpha}_{0} \diff r \right] - 
\hatDiff Z  \int \bar{\alpha}_{0} \diff r \, ,
\edm
where we have used the fact that $\Diff Z = \hatDiff Z$.
(Recall that $\Diff = \diff r \we \partial_r + \hatDiff$, 
and that the isospin harmonics are defined over $S^2$.)
Now using the expression (\ref{A13}) for $\hatDiff Z$ 
brings $\delta A^{\rm even}$ into the desired form:
\be
\delta A^{\rm even} = -\Diff \bar{\chi} + \left[
\bar{\alpha}_{1} - \frac{w}{\sqrt{2}} 
\int \bar{\alpha}_{0}
\diff r \right] \tau_{r} \hatast \diff K + \left[
\bar{\alpha}_{2} + \frac{1}{\sqrt{2}} 
\int \bar{\alpha}_{0}
\diff r \right] K \hatast \diff \tau_{r} ,
\label{A19}
\ee
where
\be
\bar{\chi} \, \equiv \, -Z  \int \bar{\alpha}_{0} \diff r .
\label{A20}
\ee
In order to separate a pure gauge 
term from $\delta H^{\rm even}$,
we use $[\taur,Z] = Y$ and $H = h \taur$ to write
\be
\delta H^{\rm even} = -[H , \bar{\chi}] + 
\bar{\gamma}_{-} X + \left[\bar{\gamma}_{+} - h
\int \bar{\alpha}_{0} \diff r \right] Y ,
\label{A21}
\ee
with $\bar{\chi}$ according to Eq. (\ref{A20}).
Hence, after a gauge transformation 
with $\bar{\chi}$, the general 
perturbations (\ref{A16}) and (\ref{A17}) 
assume the form (\ref{delta-H}) and 
(\ref{delta-A}), respectively, where
the coefficients are related as follows:
\bdm
\gamma_{-} = \bar{\gamma}_{-} , \; \; \; 
\gamma_{+} = \bar{\gamma}_{+} - h 
\int \bar{\alpha}_{0} \diff r ,
\edm
\be
\alpha_{1} = \bar{\alpha}_{1} - \frac{w}{\sqrt{2}} 
\int \bar{\alpha}_{0} \diff r , \; \; \;  
\alpha_{2} = \bar{\alpha}_{2} + \frac{1}{\sqrt{2}} 
\int \bar{\alpha}_{0} \diff r .
\label{A22}
\ee
It is clear form the above reasoning, and not hard 
to verify, that the amplitudes without bars
are gauge invariant, up to residual gauge 
transformations with 
\be
\chi_{0} = c_{1} X + c_{2} Y + c_{3} Z ,
\label{A23}
\ee
where $c_{1}$, $c_{2}$ and $c_{3}$ are arbitrary constants.
Since only the last term is relevant to the even parity 
sector, we have
$D\chi_{0}^{\rm even} = c_{3}(w \taur \hatast \diff K -
K \hatast \diff \taur)/\sqrt{2}$ and 
$[H,\chi_{0}^{\rm even}] = c_{3} h Y$.
Using this in the transformation laws (\ref{A18}) 
for the even parity perturbations
(\ref{delta-H}) and (\ref{delta-A}), we conclude that
$\gamma_{-}$ is gauge invariant, while
$\gamma_{+}$, $\alpha_{1}$ and $\alpha_{2}$ transform 
according to Eqs. (\ref{res-even}) under the residual
gauge transformations. A completely analogous 
reasoning establishes the transformation laws 
(\ref{res-odd}) for the odd parity sector.

\section{Even parity electric perturbations}
\label{AA-5}

In this Appendix we briefly show how the two coupled second
order equations (\ref{el-even}) for 
$\phi_{-}$ and $\phi_{+}$ can be translated into the 
inhomogeneous second order equation
(\ref{Sig2nd-inh}) for $\tilde{\Sigma}$, defined by 
$\tilde{\Sigma}' = h \phi'_{-}$. The procedure involves two 
integrations. The first integration is achieved by the 
observation that Eqs. (\ref{el-even}) 
can be cast into the form
\bea
\sqrt{2} \phi''_{-} & = & -2\sqrt{2}\frac{w}{r^{2}}
\left(\sqrt{2} \phi_{+} - \mu \phi_{-} \right) ,
\label{A24a}\\
\phi''_{+} \mu - \phi_{+} \mu'' & = &
-2\sqrt{2}\frac{w}{r^{2}}
\left(\sqrt{2} \phi_{+} - \mu \phi_{-} \right) ,
\label{A24b}
\eea
where we have introduced the short-hand 
$\mu \equiv w + 1/w$.
Since the RHS of the above equations are equal, 
and since both LHS are exact derivatives
[$\phi''_{+} \mu - \phi_{+} \mu'' = 
[\mu^{2}(\phi_{+}/\mu)']'$], 
an integration yields the following first 
order relation between $\phi_{-}$ and $\phi_{+}$:
\be
\frac{1}{\sqrt{2}} \left( \frac{\phi_{+}}{\mu} \right)' =
- \frac{\phi'_{-} - k_{3}}{\mu^{2}} \, ,
\label{A25}
\ee
where $k_{3}$ is an integration constant.
We now solve Eq. (\ref{A24b}) for $\phi_{+}/\mu$, perform a
derivative and use the result on the LHS of Eq. (\ref{A25}).
This yields the following second order equation 
for $\phi'_{-}$:
\be
\left(\frac{r^{2} 
\phi''_{-}}{w^{2} + 1} \right)' - 2 \phi'_{-}
=4 \, \left(\frac{w}{w^{2} + 1}\right)^{2} 
\left(\phi'_{-} - k_{3} \right) ,
\label{A26}
\ee
which shows that $\phi_{-} = const.$ is a solution of 
the system (\ref{A24a}), (\ref{A24b}).

Our aim is to integrate Eq. (\ref{A26}) once more.
In order to see that this is possible, we introduce 
the variable $\tilde{\Sigma}$ according to 
definition (\ref{def-tildesigma}), 
and note that the term in front of $(\phi'_{-} - k_{3})$
can be written in the form $[(w^{2}-1)/(w^{2}+1)]'/h$.
Hence, with
\be
\tilde{\Sigma}' \equiv h \, \phi'_{-} , \; \; \; 
a \equiv \frac{w^{2}-1}{w^{2}+1} ,
\label{A27}
\ee
Eq. (\ref{A26}) assumes the form
\be
\left[\frac{a}{h'} 
\left( \frac{\tilde{\Sigma}'}{h} \right)'
\right]' - 2 \frac{\tilde{\Sigma}'}{h} =  
\frac{a'}{h} 
\left(\frac{\tilde{\Sigma}'}{h} - k_{3} \right) .
\label{A28}
\ee
It is not hard to perform the differentiations 
and to rewrite this third order equation for
$\tilde{\Sigma}$ in the form
\bdm
\left[\left(\frac{a}{h'}\right) \tilde{\Sigma}''' + 
\left(\frac{a}{h'}\right)' \tilde{\Sigma}'' \right] - 2
\left[\left(\frac{a}{h}\right) \tilde{\Sigma}'' + 
\left(\frac{a}{h}\right)' \tilde{\Sigma}' \right] - 
\left[ 2 \tilde{\Sigma}' - k_{3} a' \right] = 0 ,
\edm
where each of the three pairs is manifestly
an exact derivative. Integrating the above expression
and multiplying the result with $h'/a$ eventually yields
\be
\tilde{\Sigma}'' - 2 \frac{h'}{h} \tilde{\Sigma}'
-2 \frac{h'}{a} \tilde{\Sigma} = -2 k_{0} \frac{h'}{a}
-k_{3} h' ,
\label{A29}
\ee
where $k_{0}$ is a further integration constant.
Since $a = r^{2} h' / (w^{2}+1)$, this
is the desired inhomogeneous second order equation
(\ref{Sig2nd-inh}). We recall that the 
four parameter family of solutions to Eq. (\ref{A29}) is
\be
\tilde{\Sigma} \, = \, \sum k_{i} \tilde{\Sigma}^{(i)} ,
\label{A30}
\ee
where the sum runs from $0$ to $3$, and where
$\tilde{\Sigma}^{(0)} = 1$ , 
$\tilde{\Sigma}^{(3)} = -h^{2}/2$,
and $\tilde{\Sigma}^{(1,2)}$ are the two 
(nontrivial) solutions to the homogeneous 
part of equation (\ref{A29}).
The four independent solutions 
(\ref{phi-0})-(\ref{phi-12}) 
to the original system (\ref{el-even}) are 
finally obtained from Eq. (\ref{def-tildesigma}).

\section{Electric contribution to the angular momentum}
\label{AA-6}

In Sect. \ref{subsection-IV-C} we have argued 
that the total angular
momentum can be expressed in terms of the
perturbation amplitudes at the origin and at infinity.
While we have established this result for the magnetic 
contribution (\ref{J-mg}), we still owe the proof of the 
formula (\ref{JEL}) for the electric part (\ref{J-el}).
In order to show that the bracket in the integrand
in Eq. (\ref{br-electric}) is an exact radial derivative,
we first perform a partial integration in 
both terms, which yields
\be
r h' \phi_{-} - \frac{r^{2}}{\sqrt{2}} \left(
\frac{w' \phi_{+}}{r} \right)' =
\left[r h \phi_{-} - \frac{r}{\sqrt{2}} w' \phi_{+} \right]'
-h \left(r \phi_{-}\right)' + \sqrt{2} w h \phi_{+} \, .
\label{A31}
\ee
In order to show that the last two terms on the RHS 
combine to an exact derivative, we use the first perturbation 
equation in (\ref{el-even}) to express $\phi_{+}$ in terms of 
$\phi_{-}$ and $\phi''_{-}$. Also using the background 
equations for $w$ and $h$, we then have
\bea
-h \left(r \phi_{-}\right)' + \sqrt{2} w h \phi_{+} & = &
-\frac{r^{2} h}{2} \, \phi''_{-} - r h \, \phi'_{-} + 
w^{2} h \, \phi_{-} 
\nonumber\\ & = &
-\left(\frac{r^{2} h}{2} \, \phi_{-}\right)'' +
\left[ w^{2} - 1 + rh \right] \phi'_{-} +
\left[2 w^{2} h + (r h)' \right] \phi
\nonumber\\ & = &
\left[ - \left(\frac{r^{2} h}{2} \, \phi_{-}\right)' +
\left( w^{2} - 1 + rh \right) \phi_{-} \right]' .
\nonumber
\eea
Using this on the RHS of Eq. (\ref{A31}) gives the desired
formula,
\be
r h' \phi_{-} - \frac{r^{2}}{\sqrt{2}} \left(
\frac{w' \phi_{+}}{r} \right)' =
-\frac{1}{2}
\left[
\left(1-w^{2}-2rh \right) \phi_{-} + r^{2} h \phi'_{-} +
\sqrt{2} w r h \phi_{+} \right]' ,
\label{A32}
\ee
which was used in Sect. \ref{subsection-IV-C} to 
establish the result (\ref{JEL}).

\end{document}